\documentclass[a4paper,11pt]{article}
\pdfoutput=1 % if your are submitting a pdflatex (i.e. if you have
\usepackage{jheppub, amsmath, amssymb,amsthm,dsfont,graphicx,mathtools,multirow,rotating,placeins,subcaption, tensor,verbatim} % for details on the use of the package, please
\usepackage[T1]{fontenc} % if needed
\usepackage[all,color]{xy}

\newcommand{\sst}[1]{{\scriptscriptstyle #1}}

\newcommand{\rep}[1]{\ensuremath{\mathbf{#1}}}

\newcommand{\s}[3]{\ensuremath{S^{\{#1, #2\}}_{#3}}}

\def\0{{\sst{(0)}}}
\def\1{{\sst{(1)}}}
\def\2{{\sst{(2)}}}
\def\3{{\sst{(3)}}}
\def\4{{\sst{(4)}}}
\def\5{{\sst{(5)}}}
\def\6{{\sst{(6)}}}
\def\7{{\sst{(7)}}}

\newcommand{\be}{\begin{equation}}
\newcommand{\ee}{\end{equation}}
\def\ba{\begin{array}}
\def\ea{\end{array}}

\newcommand\half{\tfrac{1}{2}}

\newcommand{\bea}{\begin{eqnarray}}
\newcommand{\eea}{\end{eqnarray}}

\newcommand{\jm}{j_{\max}}

\DeclareMathOperator{\tr}{tr}
\DeclareMathOperator{\diag}{diag}

\DeclareMathOperator{\SO}{SO}
\DeclareMathOperator{\USp}{Usp}
\DeclareMathOperator{\SL}{SL}
\DeclareMathOperator{\SU}{SU}
\DeclareMathOperator{\Sp}{Sp}
\DeclareMathOperator{\Spin}{Spin}
\DeclareMathOperator{\Orth}{O}

\DeclareMathOperator{\Un}{U}

\newcommand{\R}{\mathds{R}}
\newcommand{\C}{\mathds{C}}

\newcommand{\Z}{\mathds{Z}}

\newcommand{\N}{\mathcal{N}}

\title{\boldmath Twin conformal field theories}

\author[a,1]{L. Borsten,\note{Corresponding author.}}
\author[b, c, f, 2]{M.~J.  Duff, \note{The results of this paper were announced by one of us (MJD) at the Current Themes in High Energy Physics and Cosmology workshop 13-17 August 2018 Niels Bohr Institute.}}
\author[d, e]{A. Marrani}

\affiliation[a]{School of Theoretical Physics, Dublin Institute for Advanced Studies,\\
10 Burlington Road, Dublin 4, Ireland}
\affiliation[b]{Theoretical Physics, Blackett Laboratory, Imperial College London,\\
London SW7 2AZ, United Kingdom}
\affiliation[c]{Mathematical Institute, University of Oxford, Andrew Wiles Building, \\Woodstock Road, Radcliffe Observatory Quarter,
Oxford, OX2 6GG, United Kingdom}
\affiliation[d]{Museo Storico della Fisica e Centro Studi e Ricerche ``Enrico Fermi'',\\
Via Panisperna 89A, I-00184, Roma, Italy}
\affiliation[e]{Dipartimento di Fisica e Astronomia ``Galileo Galilei'', Universit\`a di Padova, and INFN, sezione di
Padova,
Via Marzolo 8, I-35131 Padova, Italy}
\affiliation[f]{Institute for Quantum Science and Engineering and Hagler Institute for Advanced Study, Texas A\&M University, College Station, TX, 77840, USA}

\emailAdd{leron@stp.dias.ie}
\emailAdd{m.duff@imperial.ac.uk}
\emailAdd{jazzphyzz@gmail.com}

\abstract{Supersymmetric theories with the same bosonic content but different fermions, aka {\it twins},  were thought to exist only for supergravity. Here we show that pairs of super conformal field theories, for example  exotic $\mathcal{N}=3$ and $\mathcal{N}=1$ theories  in $D=4$ spacetime dimensions, can also be twin. We provide evidence  from three different perspectives: (i) a twin S-fold construction, (ii) a double-copy argument and (iii) by identifying  candidate twin holographically dual  gauged supergravity theories. Furthermore, twin W-supergravity theories then follow by applying the  double-copy prescription to  exotic super conformal field theories.}

\begin{document} 
\maketitle
\flushbottom

\section{Introduction}
We argue that there exist   \emph{twin} super conformal field theories (SCFTs) having the same bosonic sectors  but distinct  supersymmetric completions. In particular, the exotic  S-fold $\mathcal{N}=3$ SCFTs \cite{Ferrara:1998zt,Aharony:2015oyb,Garcia-Etxebarria:2015wns,Aharony:2016kai} have exotic $\mathcal{N}=1$ twins based on a closely analogous construction.

In the context of supergravity, it has been long-known \cite{Gunaydin:1983rk, Dolivet:2007sz, Bianchi:2007va, Roest:2009sn,Duff:2010ss, Anastasiou:2016csv} that there exist twins     with identical bosonic  sectors, both in terms of content and couplings, but distinct degrees of supersymmetry $\N_b>\N_l$. We denote such pairs by $\{\N_b, \N_l\}$, where $b$ and $l$ refer to the `big' and `little' twin, respectively.  This is made possible by the presence  of $\N_b$ spin-3/2 gravitini in the $\N_b$-extended gravity multiplet, some of which can be traded-in for  spin-1/2 fields living in $\N_l$-extended matter multiplets. This observation would seem  to  rule  out the possibility of twin field theories with rigid supersymmetry.  However, this na\"ive obstruction is circumvented through   S-foldings that completely remove the massless states. As described in  \cite{Ferrara:2018iko}, the lowest order operator, aside from the  $12+12$ super and superconformal charges,  preserved by the S-fold projecting  onto the exotic  $\mathcal{N}=3$ SCFT is the supercurrent multiplet. It corresponds to the    $\mathcal{N}=3$ super-Weyl multiplet, which was fully  constructed in \cite{vanMuiden:2017qsh}, consisting of the massive $\Spin(3)\times \Sp(3)$ states,
\be\label{N3}
[3,2] =\mathbf{(5,1)}\oplus \mathbf{(4, 6)}\oplus \mathbf{(3, 14+1)}\oplus \mathbf{(2, 14'+6)}\oplus \mathbf{(1, 14)},
\ee
where we denote by $[\N, \jm]$ the massive $\N$-extended  long supermultiplet with  top spin $\jm$, as constructed in \cite{Ferrara:1980ra}.  We will refer to the exotic non-perturbative SCFTs of this type, obtained through an S-folding, as W-SCFTs\footnote{The nomenclature reflects: (i) the role of  Weyl multiplets \cite{Ferrara:1977ij, Bergshoeff:1980is,vanMuiden:2017qsh} in characterising the W-SCFT  spectra \cite{Ferrara:1998zt,Ferrara:2018iko}; (ii) that the product of two W-SCFTs yields a W-supergravity \cite{Ferrara:2018iko}, in analogy to the double-copy of conventional super Yang-Mills theories, which yields conventional supergravity theories.}. The S-foldings preserving $\N=3$ supersymmetry are by now reasonably well characterised and possess a number of intriguing features  \cite{Aharony:2015oyb,Garcia-Etxebarria:2015wns,Aharony:2016kai,Agarwal:2016rvx,Imamura:2016abe,Nishinaka:2016hbw, Garcia-Etxebarria:2017ffg,Bourton:2018jwb}. The focus on the $\N=3$ case is motivated, in part, by the fact that it was previously thought that for rigid supersymmetry in $D=4$ spacetime dimensions $\N=3$ necessarily  implies $\N=4$. However, the logic underlying this conclusion relies on the existence of a perturbative limit, which fails for the intrinsically non-perturbative  S-foldings. This in itself does not rule out an enhancement to $\N=4$, but the S-fold invariant operators do not fall into $\SU(4)_R$ representations, excluding this possibility \cite{Garcia-Etxebarria:2015wns}. However, there is no reason to think S-foldings are necessarily $\N=3$  and in the context of twin theories the absence of a massless sector and the presence of both spin-3/2 and spin-1/2 states   in the set of lowest dimension operators  suggests the possibility that the same bosonic content can admit different fermionic completions. Indeed, consulting \autoref{reps}, a straightforward comparison reveals that the bosonic content of \eqref{N3} is uniquely matched by 
\be
[1, 2]\oplus 14 [1, 1],
\ee
which provides the lowest order spectrum of our candidate little $\N_l=1$ twin W-SCFT, as obtained via a twin S-folding in \autoref{31}. 
Of course, this is not enough to declare them to be twin theories as, without a Lagrangian description, we have no immediate handle on the interactions. However, there are twin $\N_b=6$ and $\N_l=2$ supergravities in $D=5$, with identical bosonic sectors determined  by the common scalar coset $\SU^\star(6)/\Sp(3)$, that can be gauged with respect to the same subgroup $\SU(3)\times \Un(1)\subset \Sp(3)$.  The gauged $\N_b=6$ supergravity (or more precisely, an S-duality fibration thereof) provides the bulk holographic dual of the exotic $\N_b=3$ SCFT \cite{Ferrara:1998zt,Garcia-Etxebarria:2015wns}, while its $\N_l=2$ twin provides the candidate  bulk holographic dual of the proposed exotic $\N_l=1$ twin SCFT. Note, all  twin Poincar\'e supergravity theories can be obtained through the ``square'' or ``double-copy'' of conventional super Yang-Mills theories \cite{Anastasiou:2016csv}, as summarised in \autoref{PYR}. As one might anticipate, for each twin supergravity pair in $D=5$ there is a candidate twin pair of dual $D=4$ W-SCFTs, that admit a twin S-fold construction and   may also be  deduced through the double-copy of massive spin-1 multiplets following the procedure of \cite{Anastasiou:2016csv}, as we describe in \autoref{31} and \autoref{21}.  Since the W-SCFTs are intrinsically non-perturbative the use of ``double-copy'' here is meant rather heuristically; it is essentially an exercise in representation theory. 

Remarkably, just as the double-copy of conventional  super Yang-Mills theories gives conventional supergravity theories, it has been argued that the ``double-copy'' of W-SCFTs yields  exotic massive higher spin W-supergravity theories \cite{Ferrara:2018iko}.  The chief example is the $\N=7$ W-supergravity, which follows from the product of $\N=4$ super Yang-Mills with the $\N=3$ W-SCFT \cite{Ferrara:2018iko} and  contains a single spin-4   and 1000 spin-2 states.  Note, the existence of an $\N=7$ W-supergravity theory is the direct analog of the  existence of an $\N=3$ W-SCFT, in the sense that for  locally supersymmetric theories in $D=4$ with a perturbative limit, $\N=7$ implies $\N=8$.  Again, the loop-hole is the intrinsically non-perturbative nature of the $\N=7$ W-supergravity, which in this case can be  traced back to its $\N=3$ W-SCFT factor in the double-copy. The $\N=7$ W-supergravity has been proposed to be the effective field theory limit of a type II W-superstring theory \cite{Ferrara:2018iko}. From this perspective, the $[\N=4]\times [\N=3]$ product corresponds to the product of  $\N=4$ left-moving and   $\N=3$ right-moving fermionic strings, which follows from a  non-perturbative  string S-folding involving a T-fold and S-duality twist that acts only on the right-movers of the conventional type II string \cite{Ferrara:2018iko}.  Now, given a double-copy construction  of W-supergravities and an array of W-SCFTs with $\N=1,2,3$ one can  follow \cite{Anastasiou:2016csv} to generate candidate twin W-supergravities, as done in \autoref{Wsugra}. In this case we cannot directly appeal to AdS/CFT, so for the time-being their twinness is confined to spectra and symmetries alone.

\section{$D=4$ massive multiplets with spin $\leq 4$}

We shall need in the following all long massive supermultiplets with spins ranging from 0 to 4. For $\mathcal{N}$-extended supersymmetry, the long massive spin-$\left(\frac{\mathcal{N}}{2}+j\right)$ multiplet is obtained by tensoring the smallest long massive spin-$(\frac{\mathcal{N}}{2})$ multiplet by a spin-$j$ state \cite{Ferrara:1980ra}. This yields the  list of multiplets given \autoref{reps}. The unitary R-symmetry representations may be collected into  representations of  $\Sp(\N)$,  the automorphism algebra of the massive $\N$-extended supersymmetry algebra,  and the states are accordingly labelled by $\Spin(3)\times\Sp(\mathcal{N})$ representations.  Note the coincidences in \cite{Ferrara:2018iko} and \cite{Duff:1996qp}, which both make use of \autoref{reps}, suggesting a possible relation to bound $p$-branes.  Specifically, the $N, L, q$ multiplets of  \cite{Duff:1996qp} are related to the  ${\cal N},j_{max}$ multiplets of  \cite{Ferrara:2018iko} and \autoref{reps} by
 \be
 [{\cal N},j_{max}]=[N-q,(N-q+2L)/2].
 \ee

\section{Twin superconformal field theories}

In this section we shall construct the twin W-SCFTs in $D=4$. The $\{\N_b, \N_l\}$ twin pair can be obtained by twin S-fold operators, denoted $S^{\{\N_b, \N_l\}}_{b}$ and $S^{\{\N_b, \N_l\}}_{l}$, for the big and little twin, respectively. They each have holographic duals given by gauged twin supergravities obtained via non-perturbative projections of type IIB on $\text{AdS}_5\times S^5$. We will also show how they may deduced from using the \emph{gauge}$\times$\emph{gauge} construction of  \cite{Anastasiou:2016csv}, using a Cartan involution and $(-1)^F$.  We will review the known $\N=3$  theories, before describing the  $\{3,1\}$ and $\{2,1\}$ twins. The $\N=2,1$ W-SCFTs are new, to the best of our knowledge.

\subsection{The $\{3,1\}$ twins}
\label{31}

Before giving the S-fold construction of the twin pair, let us summarise their spectra. Let $[\N, \jm]_{B(F)}$ denote the  bosonic (fermionic) sector of  $[\N, \jm]$. 
\begin{table}
\tiny
  \[
 \hspace{-0.25in}\xymatrix{
  D=6 &&&{\begin{array}{c}\frac{\SO(5, 5)}{\SO(5)\times\SO(5)}\\[5pt]\{(2,2)\}\end{array}}\ar[r] \ar[dl]\ar@{--}@[red][d]&{\begin{array}{c}\frac{\SU^\star(4)}{\USp(2)}\\[5pt]\{(2,1), (0,1)\}\end{array}}\ar[dl]\ar@{-->}[dd]\\
    &&{\begin{array}{c}\frac{\SU^\star(4)}{\USp(2)}\\[5pt]\{(2,1), (0,1)\}\end{array}}\ar[r] \ar@{-->}[dd] &{\begin{array}{c}\frac{\Orth(1,1)\times \Sp(1)^2}{\Un(1)^2}\\[5pt]\{(1,1), (0, 1)\}\ar@{-->}@[red][d]\end{array}}\\
  D=5 &&&{\begin{array}{c}\frac{E_{6(6)}}{\USp(4)}\\[5pt]\{8\}\end{array}}\ar[r]\ar[dl] \ar@{--}@[red][d]&{\begin{array}{c}\frac{\SU^\star(6)}{\USp(3)}\\[5pt]\{6, 2\}\end{array}}\ar[dl]\ar@{-->}[dd]\\
    &&{\begin{array}{c}\frac{\SU^\star(6)}{\USp(3)}\\[5pt]\{6, 2\}\end{array}}\ar[r] \ar@{-->}[dd]&{\begin{array}{c}\frac{\SO(1,1)\times \SO(5,1)}{\USp(2)}\\[5pt]\{4, 2\}\end{array}}\ar@{-->}@[red][d]\\
  D=4 &&&{\begin{array}{c}\frac{E_{7(7)}}{\SU(8)}\\[5pt]\{8\}\end{array}}\ar[r] \ar[dl]\ar@{--}@[red][d]&{\begin{array}{c}\frac{\SO^\star(12)}{\Un(6)}\\[5pt]\{6, 2\}\end{array}}\ar[r] \ar[dl]&{\begin{array}{c}\frac{\SU(5,1)}{\Un(5)}\\[5pt]\{5, 1\}\end{array}}\ar[dl] \ar@{-->}[ddd]\\
    &&{\begin{array}{c}\frac{\SO^\star(12)}{\Un(6)}\\[5pt]\{6, 2\}\end{array}}\ar[r] \ar[dl]&{\begin{array}{c}\frac{\SU(1,1)\times\SO(6, 2)}{\Un(1)\times\Un(4)}\\[5pt]\{4, 2\}\end{array}}\ar[r] \ar[dl]\ar@{--}@[red][d]& {\begin{array}{c}\frac{\SU(3,1)}{\Un(3)}\\[5pt]\{3, 2, 1\}\end{array}}\ar[dl]\\
    &{\begin{array}{c}\frac{\SU(5,1)}{\Un(5)}\\[5pt](5, 1)\end{array}}\ar[r]  \ar@{-->}[ddd]& {\begin{array}{c}\frac{\SU(3,1)}{\Un(3)}\\[5pt]\{3, 2, 1\}\end{array}} \ar[r] &{\begin{array}{c}\frac{\SU(2,1)}{\Un(2)}\\[5pt]\{2, 1\}\end{array}}\ar@{-->}@[red][d] \\
  D=3  &&&{\begin{array}{c}\frac{E_{8(8)}}{\SO(16)}\\[5pt]\{16\}\end{array}}\ar[r] \ar[dl]&{\begin{array}{c}\frac{E_{7(-5)}}{\SO(3)\times\SO(12)}\\[5pt]\{12, 4\}\end{array}}\ar[r] \ar[dl]&{\begin{array}{c}\frac{E_{6(-14)}}{\Un(1)\times\SO(10)}\\[5pt]\{10, 2\}\end{array}}\ar[r] \ar[dl]&{\begin{array}{c}\frac{F_{4(-20)}}{\SO(9)}\\[5pt]\{9, 1\}\end{array}}\ar[dl] \\
    &&{\begin{array}{c}\frac{E_{7(-5)}}{\SO(3)\times\SO(12)}\\[5pt]\{12, 4\}\end{array}}\ar[r] \ar[dl]&{\begin{array}{c}\frac{\SO(8,4)}{\SO(8)\times\SO(4)}\\[5pt]\{8, 4\}\end{array}}\ar[r] \ar[dl]&{\begin{array}{c}\frac{\SU(4,2)}{\Un(4)\times\SU(2)}\\[5pt]\{6, 4,  2\}\end{array}}\ar[r] \ar[dl]&{\begin{array}{c}\frac{\USp(2,1)}{\USp(2)\times\SU(2)}\\[5pt]\{5, 1\}\end{array}}\ar[dl]\\
    &{\begin{array}{c}\frac{E_{6(-14)}}{\Un(1)\times\SO(10)}\\[5pt]\{10, 2\}\end{array}}\ar[r] \ar[dl]&{\begin{array}{c}\frac{\SU(4,2)}{\Un(4)\times\SU(2)}\\[5pt]\{6, 4, 2\}\end{array}}\ar[r] \ar[dl]&{\begin{array}{c}\frac{\SU(2,1)^2}{\Un(2)^2}\\[5pt]\{4, 2\}\end{array}}\ar[r] \ar[dl]&{\begin{array}{c}\frac{\SU(2,1)}{\Un(2)}\\[5pt]\{3, 1\}\end{array}}\ar[dl] \\
    {\begin{array}{c}\frac{F_{4(-20)}}{\SO(9)}\\[5pt]\{9, 1\}\end{array}} \ar[r] &{\begin{array}{c}\frac{\USp(2,1)}{\USp(2)\times\SU(2)}\\[5pt]\{5, 1\}\end{array}}\ar[r] &{\begin{array}{c}\frac{\SU(2,1)}{\Un(2)}\\[5pt]\{3, 1\}\end{array}}\ar[r] &{\begin{array}{c}\frac{\SL(2, \R)}{\SO(2)}\\[5pt]\{2, 1\}\end{array}} }
\]
\caption{{\footnotesize Pyramid of twin supergravities generated by the product of left and right super Yang-Mills theories in $D=3,4,5,6$. Each level is related by dimensional reduction as indicated by the vertical arrows. The horizontal arrows indicate consistent truncations effected by truncating the left or right Yang-Mills multiplets.  All such supergravity theories have a twin  except for the  maximal cases along the ``exceptional spine'' highlighted in red.  Note,  $D=3$ is the exception to the exceptions in that maximal $\N=16$ supergravity does have a `trivial' $\N=1$ twin, but it is not obtained from the double-copy procedure \cite{Anastasiou:2016csv}.}}\label{PYR}
\end{table}

\begin{table}
\[\scriptsize
\begin{array}{cc|lllllllllllllllllll}
j \backslash\mathcal{N} &&8&&7&&6&&5&&4&&3&&2&&1 \\[4pt]
\hline
4 			&&\mathbf{1}		&&\mathbf{1}				&&\mathbf{1}				&&\mathbf{1}				&&\mathbf{1}  			&&\mathbf{1}		&&\mathbf{1}		&&\mathbf{1}\\  
\tfrac{7}{2} 	&&\mathbf{16}		&&\mathbf{14}				&&\mathbf{12}				&&\mathbf{10}				&&\mathbf{8}			&&\mathbf{6}		&&\mathbf{4}		&&\mathbf{2}\\ 
3			&&\mathbf{119}		&&\mathbf{90+1}			&&\mathbf{65+1}			&&\mathbf{44+1}			&&\mathbf{27+1}		&&\mathbf{14+1}	&&\mathbf{5+1}	&&\mathbf{1}	\\ 
\tfrac{5}{2} 	&&\mathbf{544}		&&\mathbf{350+14} 			&&\mathbf{208+12} 			&&\mathbf{110+10}			&&\mathbf{48+8}		&&\mathbf{14'+6}	&&\mathbf{4}		&&-	\\ 
2 			&&\mathbf{1700}	&&\mathbf{910+90}			&&\mathbf{429+65+1}		&&\mathbf{165+44+1}		&&\mathbf{42+27+1}		&&\mathbf{14+1}	&&\mathbf{1}		&&-\\ 
\tfrac{3}{2} 	&&\mathbf{3808}	&&\mathbf{1638+350}		&&\mathbf{572+208+12}		&&\mathbf{132+110+10}		&&\mathbf{48+8}		&&\mathbf{6}		&&-				&&-\\ 
1 			&&\mathbf{6188}	&&\mathbf{2002+910}		&&\mathbf{429'+429+65}		&&\mathbf{165+44+1}		&&\mathbf{27+1}		&&\mathbf{1}		&&-				&&-\\ 
\tfrac{1}{2} 	&&\mathbf{7072}	&&\mathbf{1430+1638}		&&\mathbf{572+208}			&&\mathbf{110+10}			&&\mathbf{8}			&&-				&&-				&&-\\ 
0 			&&\mathbf{4862}	&&\mathbf{2002}			&&\mathbf{429}				&&\mathbf{44}				&&\mathbf{1}			&&-				&&-				&&-
\\[4pt]
\hline
\text{d.o.f}&&2^{16}&&2\times 2^{14}&&3\times 2^{12}&&4\times2^{10}&&5\times2^{8}&&6\times2^{6}&&7\times2^{4}&&8\times2^{2}\\[4pt]
\hline
\hline
\tfrac{7}{2} 	&&				&&\mathbf{1}				&&\mathbf{1}				&&\mathbf{1}				&&\mathbf{1}			&&\mathbf{1}		&&\mathbf{1}		&&\mathbf{1}\\ 
3			&&				&&\mathbf{14}				&&\mathbf{12}				&&\mathbf{10}				&&\mathbf{8}			&&\mathbf{6}		&&\mathbf{4}		&&\mathbf{2}\\ 
\tfrac{5}{2} 	&&				&&\mathbf{90}				&&\mathbf{65+1} 			&&\mathbf{44+1}			&&\mathbf{27+1}		&&\mathbf{14+1}	&&\mathbf{5+1}	&&\mathbf{1}\\ 
2 			&&				&&\mathbf{350}				&&\mathbf{208+12}			&&\mathbf{110+10}			&&\mathbf{48+8}		&&\mathbf{14'+6}	&&\mathbf{4}		&&-\\ 
\tfrac{3}{2} 	&&				&&\mathbf{910}				&&\mathbf{65+429}			&&\mathbf{165+44+1}		&&\mathbf{42+27+1}		&&\mathbf{14+1}	&&\mathbf{1}		&&-\\ 
1			&&				&&\mathbf{1638}			&&\mathbf{208+572}			&&\mathbf{132+110+10}		&&\mathbf{48+8}		&&\mathbf{6}		&&-				&&-\\ 
\tfrac{1}{2}		&&				&&\mathbf{2002}			&&\mathbf{429+429'}		&&\mathbf{165+44}			&&\mathbf{27+1}		&&\mathbf{1}		&&-				&&-\\ 
0 			&&				&&\mathbf{1430}			&&\mathbf{572}				&&\mathbf{110}				&&\mathbf{8}			&&-				&&-				&&-\\[4pt]
\hline
\text{d.o.f}&&&&2^{14} && 2\times 2^{12}&&3\times2^{10}&&4\times2^{8}&&5\times2^{6}&&6\times2^{4}&&7\times2^{2}\\[4pt]
\hline
\hline
3			&&				&&						&&\mathbf{1}				&&\mathbf{1}				&&\mathbf{1}			&&\mathbf{1}		&&\mathbf{1}		&&\mathbf{1}\\
\tfrac{5}{2}		&&				&&						&&\mathbf{12}				&&\mathbf{10} 				&&\mathbf{8} 			&&\mathbf{6}		&&\mathbf{4}		&&\mathbf{2}\\
2 			&&				&&						&&\mathbf{65}				&&\mathbf{44+1}			&&\mathbf{27+1}		&&\mathbf{14+1}	&&\mathbf{5+1}	&&\mathbf{1}\\
\tfrac{3}{2} 	&&				&&						&&\mathbf{208}				&&\mathbf{110+10}			&&\mathbf{48+8}		&&\mathbf{14'+6}	&&\mathbf{4}		&&-\\
1 			&&				&&						&&\mathbf{429}				&&\mathbf{165+44}			&&\mathbf{42+27+1}		&&\mathbf{14+1}	&&\mathbf{1}		&&-\\
\tfrac{1}{2} 	&&				&&						&&\mathbf{572}				&&\mathbf{132+110}			&&\mathbf{48+8}		&&\mathbf{6}		&&-				&&-\\
0 			&&				&&						&&\mathbf{429'}			&&\mathbf{165}				&&\mathbf{27}			&&\mathbf{1}		&&-				&&-\\[4pt]
\hline
\text{d.o.f}&&&&&&2^{12} && 2\times 2^{10}&&3\times2^{8}&&4\times2^{6}&&5\times2^{4}&&6\times2^{2}\\[4pt]
\hline
\hline
\tfrac{5}{2}		&&				&&						&&						&&\mathbf{1}				&&\mathbf{1}			&&\mathbf{1}		&&\mathbf{1}		&&\mathbf{1}\\
2			&&				&&						&&						&&\mathbf{10} 				&&\mathbf{8} 			&&\mathbf{6}		&&\mathbf{4}		&&\mathbf{2}\\
\tfrac{3}{2} 	&&				&&						&&						&&\mathbf{44}				&&\mathbf{27+1}		&&\mathbf{14+1}	&&\mathbf{5+1}	&&\mathbf{1}\\
1  			&&				&&						&&						&&\mathbf{110}				&&\mathbf{48+8}		&&\mathbf{14'+6}	&&\mathbf{4}		&&-\\
\tfrac{1}{2} 	&&				&&						&&						&&\mathbf{165}				&&\mathbf{42+27}		&&\mathbf{14+1}	&&\mathbf{1}		&&-\\
0 			&&				&&						&&						&&\mathbf{132}				&&\mathbf{48}			&&\mathbf{6}		&&-				&&-\\
\hline
\text{d.o.f}&&&&&&&&2^{10} && 2\times 2^{8}&&3\times2^{6}&&4\times2^{4}&&5\times2^{2}\\[4pt]
\hline
\hline
2			&&				&&						&&						&&						&&\mathbf{1}			&&\mathbf{1}		&&\mathbf{1}		&&\mathbf{1}\\
\tfrac{3}{2}		&&				&&						&&						&&						&&\mathbf{8} 			&&\mathbf{6}		&&\mathbf{4}		&&\mathbf{2}\\
1 			&&				&&						&&						&&						&&\mathbf{27}			&&\mathbf{14+1}	&&\mathbf{5+1}	&&\mathbf{1}\\
\tfrac{1}{2}  	&&				&&						&&						&&						&&\mathbf{48}			&&\mathbf{14'+6}	&&\mathbf{4}		&&-\\
0  			&&				&&						&&						&&						&&\mathbf{42}			&&\mathbf{14}		&&\mathbf{1}		&&-\\
\hline
\text{d.o.f}&&&&&&&&&&2^{8} && 2\times 2^{6}&&3\times2^{4}&&4\times2^{2}\\[4pt]
\hline
\hline
\tfrac{3}{2}		&&				&&						&&						&&						&&					&&\mathbf{1}		&&\mathbf{1}		&&\mathbf{1}\\
1			&&				&&						&&						&&						&&					&&\mathbf{6}		&&\mathbf{4}		&&\mathbf{2}\\
\tfrac{1}{2} 	&&				&&						&&						&&						&&					&&\mathbf{14}		&&\mathbf{5+1}	&&\mathbf{1}\\
 0 			&&				&&						&&						&&						&&					&&\mathbf{14'}		&&\mathbf{4}		&&-\\
\hline
\text{d.o.f}&&&&&&&&&& &&  2^{6}&&2\times2^{4}&&3\times2^{2}\\[4pt]
\hline
\hline
1		&&				&&						&&						&&						&&					&& 					&&\mathbf{1}		&&\mathbf{1}\\
\tfrac{1}{2}			&&				&&						&&						&&						&&					&& 			&&\mathbf{4}		&&\mathbf{2}\\
0 	&&				&&						&&						&&						&&					&& 						&&\mathbf{5}	&&\mathbf{1}\\
\hline
\text{d.o.f}&&&&&&&&&& &&  &&2^{4}&&2\times2^{2}\\[4pt]
\hline
\hline
\tfrac{1}{2}			&&				&&						&&						&&						&&					&& 			&& 		&&\mathbf{1}\\
0 	&&				&&						&&						&&						&&					&& 						&& 		&&\mathbf{2}\\
\hline
\text{d.o.f}&&&&&&&&&& &&  &&&&2^{2}\\[4pt]
\hline
\hline
\end{array}
\]
\caption{The long massive $\N$-extended spin-$\left(j_{\text{max}}\right)$ multiplets for $1/2\leq j_{\text{max}} \leq 4$ and $1\leq\N \leq 8$. The states are labelled by $\Spin(3)\times\Sp(\mathcal{N})$ representations.}\label{reps}
\end{table}
\FloatBarrier

Consider the bosonic sector of $\N=3$ spin-2 multiplet,
\be
[3,2]_B=\mathbf{(5,1)}\oplus \mathbf{(3, 14+1)}\oplus \mathbf{(1, 14)}, 
\ee
and the $\N=1$ spin-2 multiplet,
\be
[1, 2]_B=\mathbf{(5,1)}\oplus \mathbf{(3, 1)}.
\ee

Evidently, to match the bosonic content of the $\N=3$ theory we must add 14 $\N=1$ spin-1 multiplets transforming in the $\mathbf{14}$ of $\Sp(3)$, giving 
\be
[1, 2]\oplus 14 [1, 1]=\mathbf{(5,1,1)}\oplus\mathbf{(4, 2, 1)}\oplus\mathbf{(3, 1, 14+1)}\oplus\mathbf{(2, 2, 14)}
\oplus\mathbf{(1, 1, 14)}.
\ee
such that 
\be
\left([1, 2]\oplus 14 [1, 1]\right)_B = [3, 2]_B.
\ee
The $\{3,1\}$ twin multiplets, $[3,2]$ and $[1, 2]\oplus 14 [1, 1]$, can be obtained via complementary truncations of $[4,2]$. First, decompose  $[4,2]$ with respect to $\Sp(1)\times \Sp(3)\subset\Sp(4)$
\be
\begin{array}{lllllllllllllllll}
\mathbf{(5,1)}&\rightarrow&\mathbf{(5,1,1)}\\[5pt]
\mathbf{(4,8)}&\rightarrow&\mathbf{(4,2,1)}&+&\mathbf{(4,1,6)}\\[5pt]
\mathbf{(3,27)}&\rightarrow&\mathbf{(3,1,14)}&+&\mathbf{(3,2,6)}&+&\mathbf{(3,1,1)}\\[5pt]
\mathbf{(2,48)}&\rightarrow&\mathbf{(2,1,14')}&+&\mathbf{(2,2,14)}&+&\mathbf{(2,1,6)}\\[5pt]
\mathbf{(1,42)}&\rightarrow&\mathbf{(1,1,14)}&+&\mathbf{(1, 2,14')}\\[5pt]
 \end{array}
 \ee
 The $[3,2]$ multiplet is then given by truncating to the $\Sp(1)$ invariant subsector: 
\be
\begin{array}{lllllllllllllllll}
\mathbf{(5,1)}&\rightarrow&\mathbf{(5,1,1)}\\[5pt]
\mathbf{(4,8)}&\rightarrow& \mathbf{(4,1,6)}\\[5pt]
\mathbf{(3,27)}&\rightarrow&\mathbf{(3,1,14+1)}\\[5pt]
\mathbf{(2,48)}&\rightarrow&\mathbf{(2,1,14'+6)}\\[5pt]
\mathbf{(1,42)}&\rightarrow&\mathbf{(1,1,14)}\\[5pt]
 \end{array}
 \ee
 Its twin $[1, 2]\oplus 14 [1, 1]$ multiplet is given by retaining the same bosonic subector, but the complementary fermionic subsector, that is all fermions  transforming as the $\mathbf{2}$ of  $\Sp(1)$:
 \be\label{N1trunc}
\begin{array}{lllllllllllllllll}
\mathbf{(5,1)}&\rightarrow&\mathbf{(5,1,1)}\\[5pt]
\mathbf{(4,8)}&\rightarrow&\mathbf{(4,2,1)}\\[5pt]
\mathbf{(3,27)}&\rightarrow&\mathbf{(3,1,1)}&+&\mathbf{(3,1,14)}\\[5pt]
\mathbf{(2,48)}&\rightarrow&&+&\mathbf{(2,2,14)}\\[5pt]
\mathbf{(1,42)}&\rightarrow&&+&\mathbf{(1,1,14)}\\[5pt]
 \end{array}
 \ee

To summarise, decomposing under $\N=3$,
\be
[4,2] = [3,2] \oplus \rep{2}\times[3, 3/2] \quad\longrightarrow \quad [3,2] 
\ee
and the doublet of spin-3/2 multiplets are truncated out. On the other hand, decomposing under $\N=1$
\be
[4,2] = [1,2] \oplus \rep{6} \times[1, 3/2]  \oplus \rep{14} \times[1, 1] \oplus \rep{14'} \times[1, 1/2]\quad\longrightarrow \quad [1,2] \oplus \rep{14} \times[1, 1]
\ee
and the six (14) spin-3/2 (spin-1/2) multiplets are truncated out.
\subsubsection{The $\N=3$ big twin} 

Let us first recall the key features of the  $\N=3$ W-SCFTs constructed in  \cite{Garcia-Etxebarria:2015wns}.  From a field theory point of view the $\N=3$ theories are obtained by an S-fold projector, $S_{b}^{\{3,1\}}:=s_k\circ r_k$, generating a $\Z_k$ subgroup of the $\N=4$ R-symmetry and S-duality, $\Spin(6)\times\SL(2, \Z)$. The R-symmetry operator, $r_k$, 
is straight-forwardly embeded in the R-symmetry group $\Spin(6)$. Consider the   $\Z_k$ group 
\be
\Z_k\subset \Un_a(1)\times\Un_b(1)\times\Un_c(1)\subset\Spin(6).
\ee
generated by  a $(2\pi a/k, 2\pi b/k, 2\pi c/k)$ rotation on $\R^2\times\R^2\times\R^2$, for $a,b,c$ co-prime relative to $k$. Geometrically, it can be regarded as a rotation on the $\R^6$ transverse to a stack of D3-branes in $\R^{1,9}$. For  $(x,y,z)$ coordinates on $\C^3$ it is given by 
\be
(x,y,z) \mapsto (\zeta^a x, \zeta^b y, \zeta^c z), \qquad \zeta = e^{\frac{2\pi i}{k}}.
\ee
Here, $r_k$ is given by $(a,b,c)=(1,1,-1)$.   The corresponding action on the $\N=4$ supercharges is given by 
\be
r_k: \left\{\begin{array}{rrrrrrrrr} Q_{\alpha  A} &\mapsto& e^{-\frac{i2\pi\sum_l \lambda^{A}_{l}}{k}} Q_{\alpha}{}_{A}  \\[8pt]
\bar{Q}_{\dot{\alpha}}{}^{A} &\mapsto& e^{-\frac{i2\pi\sum_l \lambda_{Al}}{k}}  \bar{Q}_{\dot{\alpha}}{}^{A}
\end{array}\right.
\ee 
Here, $\alpha$ $(\dot{\alpha})$ and upper  (lower) ${A}$  are the spinor (conjugate spinor) indices of the $\mathbf{2}$ ($\mathbf{\bar{2}}$) and the $\mathbf{4}$   $(\mathbf{\bar{4}})$ representations of $\Spin(1,3)\cong \SL(2, \C)$ and  $\Spin(6)\cong\SU(4)$, respectively.  The weights of the $\mathbf{4}$   $(\mathbf{\bar{4}})$ are denoted  by $\lambda_A$  $(\lambda^A)$. Explicitly,
\be\label{N3Rsym}
r_k: \left\{\begin{array}{rrrrrrrrrrrrrr} Q_{a} &\mapsto& e^{-\frac{i\pi}{k}} Q_{a}, &\quad& Q_{4} &\mapsto& e^{\frac{i3\pi}{k}} Q_{4}  \\[8pt]
\bar{Q}^{a} &\mapsto& e^{\frac{i\pi}{k}}  \bar{Q}^{a}, &\quad& \bar{Q}^{4} &\mapsto& e^{-\frac{i3\pi}{k}}  \bar{Q}^{4}
\end{array}\right.
\ee
where $a=1,2,3$.

Consider now    S-duality (assuming a simply-laced gauge group) acting on  the coupling constant (complex structure) $\tau$ in usual fractional linear manner\footnote{It is the projective $\text{PSL}(2, \Z)$ that acts faithfully on  the upper-half plane, but since we will consider the S-duality action on the fermionic supercharges its double-cover $\text{SL}(2, \Z)$ is required.},
\be
\tau\mapsto \frac{a\tau+b}{c\tau+d},\qquad \left(\begin{array}{cc}a&b\\c&d\end{array}\right)\in \text{SL}(2, \Z).
\ee
The corresponding action on the  supercharges is given by, 
\be
Q_A \mapsto \sqrt{\frac{c\tau+d}{|c\tau+d|}} Q_A,\quad \bar{Q}^A \mapsto \sqrt{\frac{|c\tau+d|}{c\tau+d}} \bar{Q}^A,
\ee
where the central charge picks up a factor of $\frac{|c\tau+d|}{c\tau+d}$ under S-duality and the presence of the squareroot implies that the supercharges in the   double-cover.  
 Note, S-duality is only a symmetry (as opposed to a duality) if $\tau$  is preserved. This only happens for particular subgroups $\Gamma\subset \SL(2, \Z)$ corresponding to certain values of $\tau$,    in which case  $Q_A\mapsto \exp[i\pi/k]Q_A$ 
for specific values of $k$ depending on $\tau$, as summarised here: 
\be\label{tauval}
\begin{array}{c|ccccccc}
\hline
\hline
\Gamma &\Z_2&\Z_3&\Z_4&\Z_2\times\Z_3\\[5pt]
\text{Generator}& \left(\begin{array}{cc}-1&0\\0&-1\end{array}\right)& \left(\begin{array}{cc}-1&1\\-1&0\end{array}\right) & \left(\begin{array}{cc}0&-1\\1&0\end{array}\right)& \left(\begin{array}{cc}1&-1\\1&0\end{array}\right)\\[15pt]
\hline
\tau& \text{any}&e^{\frac{i\pi}{3}}& i & e^{\frac{i\pi}{3}}\\
\hline
k&2&3&4&6\\
\hline
\hline
\end{array}
\ee
 Corresponding to the $\Z_k$ R-symmetry operator \eqref{N3Rsym}  consider  a  $\mathds{Z}_k\subset \SL(2, \mathds{Z})$ S-duality subgroup generated by $s_k$, 
\be\label{sk}
s_k: \left\{\begin{array}{rrrrrrrrr} Q_{A} &\mapsto& e^{\frac{i\pi}{k}} Q_{A}  \\[8pt]
\bar{Q}^{A} &\mapsto& e^{-\frac{i\pi}{k}}  \bar{Q}^{A}
\end{array}\right.
\ee
The composite $s_k\circ r_k$ action is given by 
\be\label{b31twin}
S^{\{3,1\}}_{b}: \left\{\begin{array}{rrrrrrrrrrrrrr} Q_{a} &\mapsto&  Q_{a}, &\quad& Q_{4} &\mapsto& e^{\frac{4\pi i}{k}}  Q_{4};  \\[8pt]
\bar{Q}^{a} &\mapsto&   \bar{Q}^{a}, &\quad&
\bar{Q}^{4} &\mapsto&  e^{-\frac{4\pi i}{k}}  \bar{Q}^{4}.
\end{array}\right.
\ee
For $k=2$, corresponding to the usual orientifold case,  we see all 16 supercharges are preserved. On the other hand for $k>2$, only the 12 supercharges $Q_{a}, \bar{Q}^{a}$ are left invariant, reducing the $\N=4$ algebra to the $\N=3$  algebra, while the $\SU(4)_R$ R-symmetry is broken to $\Un(3)_R$.  For $k=2$, $\tau$ can take arbitrary values and there is a perturbative limit, as expected for the standard orientifold. For $k>2$, $\tau$ has a fixed value of order one and the S-fold is intrinsically non-perturbative.

\begin{table}
\[
\begin{array}{ccccccccc}
\hline
\hline
&&\SU(3)&&\Un(1)_R&& \s31l\\[4pt]
\hline
F^+&&\rep{1}&&0&& 1\\[4pt]
\hline
\lambda^{a}&&\rep{3}&&\hspace{-0.3cm}-1&& 1\\[4pt]
\lambda^{4}&&\rep{1}&&3&&\hspace{-0.3cm}-1\\[4pt]
\hline
\phi^{a4}&&\rep{3}&&2&&\hspace{-0.3cm}-1\\[4pt]
\phi^{ab}&&\rep{\bar{3}}&&\hspace{-0.3cm}-2&&1\\[4pt]
\hline
\hline
\end{array}
\]
\caption{The charges carried by the component fields of the $\N=4$ super Yang-Mills multiplet under the $\s31b$ S-fold operator (in units of $2\pi/k$) and the invariant $\SU(3)\times \Un(1)_R\subset \SU(4)$  R-symmetry subgroup. }\label{N3reps}
\end{table}

The entire $\N=4$ vector multiplet transforms non-trivially under $\s31b$, as summarised in \autoref{N3reps}. The spectrum is truncated to the $\s31b$-invariant  subsector. The lowest  $S_k$-invariant  operator, which has scaling dimension two, is the $\N=3$ supercurrent multiplet, which can be written 
\be
J_{a}{}^{b} = \tr \left(V_a\overline{V}^b-\frac{1}{3}\delta_{a}{}^{b}V_c \overline{V}^c\right)
\ee
where the $\Un(3)$ triplet $V_a$ is the  $\N=3$ spin-1 on-shell superfield \cite{Siegel:1980bp, Howe:1981qj}. The physical components in terms of $\Spin(3)\times \Un(3)_R$ representations are given by 
\be\label{31breps}
\begin{split}
[3,2] &=~~\mathbf{(5,1)}\\
&\phantom{=} \oplus {(\mathbf{4}, \mathbf{3}_{1}+\mathbf{\bar{3}}_{-1})} \\
&\phantom{=}\oplus (\mathbf{3}, \mathbf{1}_{0} + \mathbf{3}_{-2}+\mathbf{\bar{3}}_{2}+\mathbf{8}_{0})\\
&\phantom{=}\oplus \mathbf{(2,  \mathbf{1}_{3}+\mathbf{1}_{-3}+\mathbf{6}_{-1}+\mathbf{\bar{6}}_{1}+\mathbf{3}_{1}+\mathbf{\bar{3}}_{-1})}\\ &\phantom{=} \oplus \mathbf{(1, \mathbf{3}_{-2}+\mathbf{\bar{3}}_{2}+\mathbf{8}_{0})}\\
&=\mathbf{(5,1)}\oplus \mathbf{(4, 6)}\oplus \mathbf{(3, 14+1)}\oplus \mathbf{(2, 14'+6)}\oplus \mathbf{(1, 14)}
\end{split}
\ee
where in the last line we have collected the  $\Un(3)$ representations into $\Sp(3)$ representations corresponding to the automorphism algebra of the massive $\N=3$ supersymmetry algebra.

The $\N=3$ theories have string/M-theory embeddings  that can be approached from a number of perspectives \cite{Garcia-Etxebarria:2015wns}. For example, D3-branes probing singularities  in F-theory  on an Abelian orbifold,
\be
\underline{\R^{1,3}}\times (\C^3\times T^2)/\Z_k,
\ee  
where the underline denotes the D3-brane world-volume directions. This corresponds to a limit of M2-branes in M-theory on $\underline{\R^{1,2}}\times (\C^3\times T^2)/\Z_k$. The complex structure of the F-theory $T^2$ is the coupling constant of  the world-volume theory of the D3-branes. For $(x,y,z, u)$ coordinates on $\C^4$,  locally equivalent to  the F-theory $\C^3\times T^2$,  consider the $\Z_k$ group generated by  
\be\label{Zk}
\sigma: (x,y,z, u) \mapsto (\zeta^a x, \zeta^b y, \zeta^c z, \zeta^d u)
\ee
where $\zeta$ is a primitive $k^{\text{th}}$ root of unity. The singularities are isolated if and only if the weights $(a,b,c,d)$ are all relatively prime to $k$.  This action  is embedded in the R-symmetry and S-duality groups, 
\be
\Z_k\subset \Un_a(1)\times\Un_b(1)\times\Un_c(1)\times\Un_d(1)\subset\Spin(6)\times \SL(2, \Z),
\ee
 through
 \be
 \sigma \mapsto \diag(R_a, R_b, R_c) \otimes R_d
 \ee
 where $R_a$ is a rotation by $2\pi a/k$ on $\R^2\cong \C \subset \C^4$.   The corresponding action on the $(\rep{4}, \rep{2})$ of $\Spin(6)\times \SL(2, \Z)$ is given (in our conventions\footnote{The weights of the $\rep{4}$ are given by $(\pm \half,\pm \half,\pm \half)$ with an even number of negative signs.}) by,
 \be
\diag(\zeta^{\frac{a+b+c+d}{2}}, \zeta^{\frac{a-b-c+d}{2}}, \zeta^{\frac{-a+b-c+d}{2}}, \zeta^{\frac{-a-b+c+d}{2}}),
 \ee
where the $S^{3,1}_{b}$ action  given in \eqref{b31twin}, corresponds to $(a,b,c,d)=(1, 1, -1, -1)$,
\be\label{31b}
\sigma: (x,y,z, u) \mapsto (\zeta x, \zeta y, \zeta^{-1} z, \zeta^{-1} u).
\ee
 The singularities are $\mathds{Q}$-factorial (being quotient) Gorenstein terminal\footnote{A singular variety is said to be Gorenstein if its canonical bundle (which may only be a  coherent sheaf) is a line bundle. The quotients $\C^4/\Z_k$ are Gorenstein terminal if and only if there is  a generator with weights  given  (up to permutations) by $(1, -1, a, -a)$ for $\gcd(a, k)=1$ \cite{morrison1984terminal}.} and therefore do not admit any crepent resolution. The  $\SL(2, \Z)$ action is an involution of the torus only for $k=2,3,4,6$, hence the restriction. Since the complex structure in the F-theory limit corresponds to the axion-dilaton, for $k>2$ this is non-pertubative in  $D=10$.

\subsubsection{The $\N=1$ little twin}\label{21little}  

 Let us now introduce the little twin S-fold operator $S^{\{3, 1\}}_{l}:=s_k\circ r_k$. The R-symmetry action is again given by, 
\be
r_k: \left\{\begin{array}{rrrrrrrrrrrrrr} Q_{a} &\mapsto& e^{-\frac{i\pi}{k}} Q_{a}, &\quad& Q_{4} &\mapsto& e^{\frac{i3\pi}{k}} Q_{4}  \\[8pt]
\bar{Q}^{a} &\mapsto& e^{\frac{i\pi}{k}}  \bar{Q}^{a}, &\quad& \bar{Q}^{4} &\mapsto& e^{-\frac{i3\pi}{k}}  \bar{Q}^{4}
\end{array}\right.
\ee
The S-duality operator, on the other hand, is now given by,
\be\label{sk}
s_k: \left\{\begin{array}{rrrrrrrrr} Q_{A} &\mapsto& e^{-\frac{i3\pi}{k}} Q_{A};  \\[8pt]
\bar{Q}^{A} &\mapsto& e^{\frac{i3\pi}{k}}  \bar{Q}^{A}. 
\end{array}\right.
\ee
Hence, the composite action is given by
\be\label{l31twin}
S^{\{3,1\}}_{l}: \left\{\begin{array}{rrrrrrrrrrrrrr} Q_{a} &\mapsto& e^{\frac{-i4\pi}{k}}  Q_{a}, &\quad& Q_{4} &\mapsto&  Q_{4};  \\[8pt]
\bar{Q}^{a} &\mapsto& e^{\frac{i4\pi}{k}}  \bar{Q}^{a}, &\quad&
\bar{Q}^{4} &\mapsto&   \bar{Q}^{4},
\end{array}\right.
\ee
and we observe that for $k>2$ only four of the superchrages are left invariant. For $k=2$ all 16 charges survive as before.
The R-symmetry is broken to $\SU(3)\times \Un(1)_R$, but for $k>2$ the remnant $\SU(3)$ is now a flavour symmetry rather than an R-symmetry, since only four supercharges are left invariant, reducing the $\N=4$ algebra to the $\N=1$  algebra.  Note,  this is the unique (up to trivial automorphisms) $\N=1$ projection preserving an $\SU(3)$ subgroup of the  $\N=4$ R-symmetry.

\begin{table}
\[
\begin{array}{ccccccccc}
\hline
\hline
&&\SU(3)&&\Un(1)_R&& \s31l\\[4pt]
\hline
F^+&&\rep{1}&&0&& \hspace{-0.3cm}-3\\[4pt]
\hline
\lambda^{a}&&\rep{3}&&\hspace{-0.3cm}-1&& \hspace{-0.3cm}-1\\[4pt]
\lambda^{4}&&\rep{1}&&3&&\hspace{-0.3cm}-3\\[4pt]
\hline
\phi^{a4}&&\rep{3}&&2&&\hspace{-0.3cm}-1\\[4pt]
\phi^{ab}&&\rep{\bar{3}}&&\hspace{-0.3cm}-2&&1\\[4pt]
\hline
\hline
\end{array}
\]
\caption{The charges carried by the component fields (in terms of on-shell field strengths) of the $\N=4$ super Yang-Mills multiplet under the $\s31l$ S-fold operator (in units of $2\pi/k$) and the invariant $\SU(3)\times \Un(1)_R\subset \SU(4)$  flavour/R-symmetry subgroup. Note, for $k=3$ both $F$ and $\lambda^4$ are $\s31l$-invariant and we   restrict to $k=4$ to obtain the little twin.}\label{N1reps}
\end{table}

Again, for $k\not=3$ the entire $\N=4$ vector multiplet transforms non-trivially under $\s31l$, as summarised in \autoref{N1reps}. To obtain the desired  little twin one must set $k=4$ (other values give further truncations). Using the $\s31l$ charges it is straightforward to deduce the quadratic $\s31l$-invariant operators and, through their $\SU(3)\times\Un(1)_R$ representations, to collect them into massive long $\N=1$ supermultiplets. In term of the on-shell superfields of \cite{Siegel:1980bp, Howe:1981qj} we obtain a single spin-2
and 14 spin-1 supercurrents. This can  be deduced directly from the $\s31l$-invariant truncation of the $\N=4$ supercurrent, 
\be
J_{AB,CD} = V_{AB}V_{CD}-\frac{1}{12}\epsilon_{ABCD}\bar{V}^{EF} {V}_{EF}, \qquad \bar{V}^{AB}=\frac{1}{12}\epsilon^{ABCD}V_{CD}.
\ee The explicit projection in terms of off-shell component fields is given in \autoref{field-proj}.

The   spin-2 supercurrent  corresponds to  the massive $\mathcal{N}=1$ super-Weyl multiplet, see \autoref{reps}, which consists of the massive $\Spin(3)\times \Un(1)_R$ states
\be
\begin{split}\label{J1}
[1,2] &=\mathbf{5_0}+ {\mathbf{4}_{1}+\mathbf{4}_{-1}} + \mathbf{3}_{0} \\
&=\mathbf{(5,1)}+ \mathbf{(4, 2)}+ \mathbf{(3,1)}
\end{split}
\ee
where in the last line we have collected the  $\Un(1)$ representations into $\Sp(1)$ representations corresponding to the automorphism algebra of the massive $\N=1$ supersymmetry algebra.

The 14 spin-1  supercurrents, $ J_a,  J^a, J_{a}{}^{b}$ transform as the $\rep{3_{-2}, \bar{3}_{2}, 8_{0}}$ of the global $\Un(3)$ and can be put in a $\rep{14}$ of $\Sp(3)$,
\be
\begin{array}{cccc}
\Sp(3)&\supset &\Un(3)\\
\rep{14}&\longrightarrow &\rep{3_{-2} + \bar{3}_{2} + 8_{0}},
\end{array}
\ee
although in this case $\Sp(3)$ is a flavour symmetry, rather than the supersymmetry algebra automorphism group.  The   spin-1 supercurrents correspond to  the massive long $\mathcal{N}=1$ spin-1 multiplet, see \autoref{reps}, which consists of the massive $\Spin(3)\times \Un(1)_R$ states
\be
\begin{split}
 [1,1] &=\mathbf{3_0}+ \mathbf{2}_{1}+\mathbf{2}_{-1}+\mathbf{1}_{0}\\
&=\mathbf{(3,1)}+ \mathbf{(2, 2)}+ \mathbf{(1,1)}
\end{split}
\ee
where  we have collected the  $\Un(1)_R$ representations into $\Sp(1)$ representations corresponding to the automorphism algebra of the massive $\N=1$ supersymmetry algebra.   Including the flavour symmetry we have 
\be
14\times [1,1] 
=\mathbf{(3,1, 14)}+ \mathbf{(2, 2, 14)}+ \mathbf{(1,1, 14)}
\ee
which, with \eqref{J1}, reproduces the truncation given in \eqref{N1trunc}.

Note, geometrically the $\s31l$ projection corresponds to     $(\C^3\times T^2)/\Z_k$,
where   the $\Z_k$ action  is given by 
\be\label{31l}
 (x,y,z, u) \mapsto (\zeta x, \zeta y, \zeta^{-1} z, \zeta^{3} u).
\ee
For $k=2$ all supercharges are left invariant and we return to the orientifold case. For, $k=3,6$ the  singularities are not isolated. So, for $\N=1$ supersymmetry and  isolated singularities we must restrict to $k=4$, in which case  \eqref{31l} reduces to the $\N=3$ quotient given in \eqref{31b}. However, since the supercharges transform in the double-cover of the duality group they must be distinguished. For $k=2,3,6$ the singularities are  terminal\footnote{Isolated quotient singularities $\C^4/\Z^k$ are terminal if and only if, $s_p>k$ for $p=1,2,\ldots k-1$, where $s_p:=\langle pa \rangle +\langle pb \rangle+\langle pc \rangle+\langle pd \rangle$  and $\langle x \rangle$ is the unique integer in $\{0,1,2,\ldots k-1\}$ congruent to $x$ mod $k$ \cite{Reid:1979}.}  (but not Gorenstein) and therefore do not admit any crepent resolution. In fact, there is no isolated quotient  singularity   with $\N<3$ supersymmetry  that is Gorenstein terminal for any $k=2,3,4,6$. The actual string/F-theory embedding is rather more subtle; we shall return to this question in future work.

\subsubsection{Dual  supergravity theories}\label{31dual}

Ungauged  $D=5, \N_b=6$ supergravity has a twin given by the $\N_l=2$  quaternionic magic  supergravity, which is coupled to 14 vector multiplets and is based on the Jordan algebra of $3\times 3$ quaternionic Hermitian matrices, $\mathfrak{J}^{\mathds{H}}_{3}$ \cite{Gunaydin:1983rk, Gunaydin:1984ak}. The   bosonic sectors of the twins are  determined  by the common scalar coset $\SU^\star(6)/\Sp(3)$, where $\SU^\star(6)$ is the reduced structure group of $\mathfrak{J}^{\mathds{H}}_{3}$. This is the $D=5$ analog of the ungauged $D=4, \{6,2\}$ twins with common coset  $\SO^\star(12)/\Un(6)$. In this case, there are twins gaugings with the same $\Un(4)$ gauge group \cite{Roest:2009sn, Andrianopoli:2008ea}. The gauged $\N=6$ theory corresponds to the low energy limit of type II strings on a specific $\text{AdS}_4\times \C\mathds{P}^3$ geometry, but cannot viewed as spontaneously broken phase of a gauged $\N=8$ supergravity \cite{Andrianopoli:2008ea}. The same applies to the gauged $\N=2$ twin. Rather, they are consistent truncations of the $\SO(8)$ gauged $\N=8$ theory.

An analogous discussion  applies to the $D=5$, $\{6,2\}$ supergravity twins relevant here.  Indeed, one can consistently truncate from $\SU(4)$ gauged $\N=8$ supergravity on an $\text{AdS}_5$ background (geometrically obtained from type IIB supergravity  on $S^5$)  to both an $\Un(3)\subset \SU(4)$ gauged $\N=6$ supergravity or an $\Un(1)_R\times \Un(3)\subset \SU(4)$  gauged $\N=2$ supergravity coupled to eight vector multiplets and $3+3$ ``self-dual'' tensor multiplets, transforming as the $\rep{8}$ and $\rep{3+\bar{3}}$ of $\SU(3)\subset \SU(2,2|1)\times \SU(3)\subset \SU(2,2|4)$ respectively  \cite{Gunaydin:1985cu}. Note, the $\N=2$ multiplet structure is precisely reflected by the candidate little twin $\N=1$  W-SCFT dual obtained from the S-folding in \autoref{21little}, cf.~\autoref{field-proj}. Moreover, the $\N=6$ and $\N=2$ truncations correspond to  `twin gaugings' of  the twin  $\N=6$ and magic $\N=2$ Poincar\'e supergravity theories  \cite{Gunaydin:1984ak, Gunaydin:1985cu}, appearing in \autoref{PYR}, which both have scalar coset $\SU^\star(6)/\Sp(3)$. 
 As discussed in \cite{Ferrara:1998zt}, the  $\N=6$ truncation  should have a   dual theory with superconformal group $\SU(2,2|3)$: the  $\Un_1(1)$ projector used in \cite{Ferrara:1998zt}, which is a linear combination of a $\Un(1)$  R-symmetry and a (discretized) $\Un(1)$ S-duality, to effect the truncation, eliminates all states in the dual theory not corresponding to $\N=3$ operators. It corresponds directly to the dual big twin $\N=3$ W-SCFT S-fold.  Note, there is no conventional geometric symmetry that can effect this truncation and the use of S-duality (which for $k>2$ fixes the  string coupling to order one) makes it intrinsically non-perturbative \cite{Ferrara:1998zt, Garcia-Etxebarria:2015wns}. 
This  provides the holographic dual  of the  $\N=3$ W-SCFT \cite{Garcia-Etxebarria:2015wns}. Specifically, it is given by type IIB on $\text{AdS}_5\times S^5/\Z_k$ with a non-trivial S-duality bundle over the internal space. The corresponding F-theory construction is given by compactifiying on $\text{AdS}_5\times (S^5 \times T^2)/\Z_k$. The analogous consistent truncation to the little $\N=2$ gauged supergravity should be effected  by the same procedure used in \cite{Ferrara:1998zt} for $\N=6$, but with the little twin S-duality $\Un(1)$ rotation (i.e.~it is shifted as for the little twin W-SCFT, $\exp i\theta \pi\rightarrow \exp -3 i\theta \pi$), and should have a   dual theory with superconformal group $\SU(2,2|1)$ that  corresponds to the little $\N=1$ W-SCFT. Again, there is no conventional geometric symmetry that can effect this truncation and the use of S-duality  makes it intrinsically non-perturbative.  The complete (non)-geometric  picture will be developed in future work.
 
\subsubsection{The double-copy construction}\label{dctwin} 

Following \cite{Anastasiou:2016csv} the $\{3,1\}$ twin theories  may be generated by considering the product of Left and Right $\N=2, 1, 0$ spin-1 theories.  Assume the spin-1 theories have gauge groups $G$ and $\tilde{G}$, with Lie algebras $ \mathfrak{g}$ and $\tilde{\mathfrak{g}}$, and are valued in  the respective adjoint representations, $A$ and $\tilde{A}$. The $\N=4$ parent theory is given by,
\be
  [2,1]^A \otimes [2,1]^{\tilde{A}} =[4,2].
\ee
Consider a subgroup $G_0\subset G$ corresponding to the positive eigenspace subspace of a Cartan involution $\theta: \mathfrak{g}\rightarrow \mathfrak{g}$. The  adjoint representation decomposes as $A = A_0\oplus \rho$, where  $\rho$ is a (not necessarily irreducible) representation of $G_0$.  
To obtain the $\N=3$ twin, first decompose the Left factor  into $\N=1$ multiplets,
\be
\left([1,1]^{A} \oplus \rep{2} [1,\tfrac{1}{2}]^{A} \right) \otimes [2,1]^{\tilde{A}} =[4, 2],
\ee
where the multiplicities are given as representations of $\Sp(1)_F$ in $\Sp(1)_R\times\Sp(1)_F\subset\Sp(2)_R$.
Then let $\sigma:=(-1)^F\circ \theta$, where $(-1)^F[\N, j] = (-1)^{2j}[\N, j]$, and truncate to the $\sigma$-invariant sector of the Left factor
\be\label{DCN3}
   \left([1,1]^{A_0}  \oplus \rep{2} [1,\tfrac{1}{2}]^{\rho} \right) \otimes [2,1]^{\tilde{A}} = [3,2],
\ee
 where we have used the rule that adjoint and non-adjoint representations do not talk to one another in the double-copy and the remaining  total global symmetry is $\Sp(3)_R\times\Sp(1)_F$. 

To then obtain the $\N=1$ twin,   decompose the Right factor into $\N=0$ multiplets and truncate to the $\tilde{\sigma}:=(-1)^F\circ \tilde{\theta}$ invariant sector 
\be\label{DCN1}
   \left([1,1]^{A_0}  \oplus \rep{2} [1,\tfrac{1}{2}]^{\rho} \right) \otimes \left([0,1]^{\tilde{A}_0} \oplus \rep{4} [0,\tfrac{1}{2}]^{\tilde{\rho}} \oplus \rep{5} [0,0] ^{\tilde{A}_0} \right)
\ee
where the right multiplicities are given as representations of $\Sp(2)_{\tilde{F}}\subseteq\Sp(2)_{\tilde{R}}$. This yields 
\be
\rep{(1,1)}[1,2]+ (\rep{(1,1)+(2,4)+(1,5)}) [1,1],
\ee
where the  multiplicities are given as representations of $\Sp(1)_F\times \Sp(2)_{\tilde{F}}$ and can be collected into irreducible $\Sp(3)_F$ representations
\be
[1,2]+ \rep{14} [1,1],
\ee
so that the total global symmetry is $\Sp(1)_R\times\Sp(3)_F$. Hence, the spectra and symmetries match those of the big $\N=3$ twin. 

Note, the conventional Bern-Carrasco-Johansson (BCJ) double-copy \cite{Bern:2008qj, Bern:2010ue, Bern:2010yg}  takes gauge theories into gravitational theories, whereas here we are generating the spectra and symmetries of \emph{non-gravitational} W-SCFTs from the product of spin-1 SCFT ``matter'' multiplets.  This is directly analogous to the BCJ double-copy of, for example, $\N=2$ hyper multiplet amplitudes, which generate the amplitudes of $\N=4$ Yang-Mills. However, for the hyper  multiplets to have a local symmetry they must come coupled to an $\N=2$ Yang-Mills multiplet, which will generate the $\N=4$ gravitational sector when included in the double-copy. So the $\N=4$ Yang-Mills amplitudes generated by the hypers  must be regarded as a subsector of the full double-copy theory including the gravitational degrees of freedom. It is tempting to apply the same logic in the present W-SCFT case: the product of the ``matter'' multiplets, given in \eqref{DCN3} and \eqref{DCN1}, yields the non-gravitational twin W-SCFTs, but if they are to have local symmetries the ``matter'' multiplets entering in the Left and Right factors must themselves come coupled to W-SCFTs, which when included in the product will yield the gravitational sector in terms of W-supergravities, as described in \cite{Ferrara:2018iko} and \autoref{Wsugra}. So, in the end, we expect our double-copy constructed W-SCFTs to come coupled to W-supergravities. 

Of course, this remains rather heuristic since we dealing with non-Lagrangian theories with no perturbative limit, although using the field-theoretic approach of \cite{Anastasiou:2014qba, Anastasiou:2016csv,Anastasiou:2017nsz,  Anastasiou:2018rdx} the spectra and local/global symmetries can be determined from the product, even in the absence of a complete understanding of the factors.  It may be possible to make further progress by studying the possible rational superconformal invariants, but we leave this for future work.

\subsection{The  $\left\{ 2,1\right\}$ twins}
\label{21}
Before giving the S-fold construction of the twin pair, let us summarise their spectra. Consider the $\mathcal{N}=2$ and $\mathcal{N}=1$ super-Weyl multipets
\begin{eqnarray}
\lbrack 2,2] &=&\left( \mathbf{5},\mathbf{1}\right) +\left( \mathbf{4},%
\mathbf{4}\right) +\left( \mathbf{3},\mathbf{5+1}\right) +\left( \mathbf{2},%
\mathbf{4}\right) +\left( \mathbf{1},\mathbf{1}\right) ; \\
\lbrack 1,2] &=&\left( \mathbf{5},\mathbf{1}\right) +\left( \mathbf{4},%
\mathbf{2}\right) +\left( \mathbf{3},\mathbf{1}\right),
\end{eqnarray}%
which are covariant under $\Spin(3)\times \Sp(2)$ and $\Spin(3)\times \Sp(1)$, respectively.
Consequently, in order to equate their bosonic sectors, one must add at least
 five $[1,1]$ multiplets to the $\mathcal{N}=1$ theory, transforming in the $%
\mathbf{5}$ of $\Sp(2)$, giving the  $\Spin(3)\times \Sp(1)\times
\Sp(2)$-covariant result,%
\begin{eqnarray}
\lbrack 2,2]_{B} &=&\left( \mathbf{5},\mathbf{1,1}\right) +\left( \mathbf{3},%
\mathbf{1},\mathbf{5+1}\right) +\left( \mathbf{1},\mathbf{1,1}\right) ; \\
\left( \lbrack 1,2]\oplus \mathbf{5}[1,1]\right) _{B} &=&\left( \mathbf{5},%
\mathbf{1,1}\right) +\left( \mathbf{3},\mathbf{1,5+1}\right) +\left( \mathbf{%
1},\mathbf{1},\mathbf{5}\right) .
\end{eqnarray}%
The unique, minimal matching of the   bosonic sectors is then given by adding one $[2,1]$
multiplet on the $\mathcal{N}=2$ side, and a further $[1,1]$ multiplet on the $%
\mathcal{N}=1$ side:%
\begin{eqnarray}
\left( \lbrack 2,2]\oplus \lbrack 2,1]\right) _{B} &=&\left( [1,2]\oplus
\left( \mathbf{5+1}\right) [1,1]\right) _{B}  \nonumber \\
&=&\left( \mathbf{5},\mathbf{1,1}\right) +\left( \mathbf{3},\mathbf{1,5+1+1}%
\right) +\left( \mathbf{1},\mathbf{1},\mathbf{5+1}\right) .
\end{eqnarray}%
Thus, the $\left\{ 2,1\right\} $ twin W-SCFT pair in $D=4$ is given
by the $\mathcal{N}=2$ W-SCFT with $[2,2]\oplus \lbrack 2,1]$ and the $%
\mathcal{N}=1$ W-SCFT with $[1,2]\oplus \left( \mathbf{5+1}\right) [1,1]$,
where $\mathbf{5+1}$ is a reducible representation of $\Sp(2)$.

The $\left\{ 2,1\right\} $ twin multiplets can be obtained via complementary
truncations of $[3,2]$. First, decompose $[3,2]$ under $\Sp(1)\times
\Sp(2)\subset \Sp(3)$:%
\begin{equation}
\begin{array}{lllllllllll}
\left( \mathbf{5},\mathbf{1}\right)  & \rightarrow  & \left( \mathbf{5},
\mathbf{1,1}\right)  &  &  \\
\left( \mathbf{4},\mathbf{6}\right)  & \rightarrow  & \left( \mathbf{4},\mathbf{1,4}\right)  & + &  \left( \mathbf{4}, \mathbf{2,1}\right)  &  \\
\left( \mathbf{3},\mathbf{14}+\mathbf{1}\right)  & \rightarrow  & \left( \mathbf{3}, \mathbf{1,5}\right)&+& \left(\mathbf{3},\mathbf{2,4}\right) &+ & \left( \mathbf{3}, \mathbf{1,1}\right)   &+ & \left( \mathbf{3}, \mathbf{1,1}\right) \\
\left( \mathbf{2},\mathbf{14}^{\prime }+\mathbf{6}\right)  & \rightarrow  &  \left( \mathbf{2},\mathbf{1,4}\right)  &+ & \left( \mathbf{2}, \mathbf{2, 5}\right)  &+ &\left( \mathbf{2},\mathbf{1,4}\right)  &+ &    \left( \mathbf{2}, \mathbf{2, 1}\right)   \\
\left( \mathbf{1},\mathbf{14}\right)  & \rightarrow  &  \left( \mathbf{1},\mathbf{1,5}\right) &+ & \left(\mathbf{1},\mathbf{2,4}\right)&+ &  \left( \mathbf{1}, \mathbf{1,1}\right).
\end{array}
\end{equation}
The truncation to the $\Sp(1)$-invariant subsector yields the $\mathcal{N}=2$
twin $[2,2]\oplus \lbrack 2,1]$:%
\begin{equation}\label{Sp1trunc}
\begin{array}{lllllllllll}
\left( \mathbf{5},\mathbf{1}\right)  & \rightarrow  & \left( \mathbf{5},
\mathbf{1,1}\right)  &  &  \\
\left( \mathbf{4},\mathbf{6}\right)  & \rightarrow  & \left( \mathbf{4},\mathbf{1,4}\right)     \\
\left( \mathbf{3},\mathbf{14}+\mathbf{1}\right)  & \rightarrow  & \left( \mathbf{3}, \mathbf{1,5+1}\right)   &+ & \left( \mathbf{3}, \mathbf{1,1}\right) \\
\left( \mathbf{2},\mathbf{14}^{\prime }+\mathbf{6}\right)  & \rightarrow  &  \left( \mathbf{2},\mathbf{1,4}\right)   &+ &\left( \mathbf{2},\mathbf{1,4}\right)    \\
\left( \mathbf{1},\mathbf{14}\right)  & \rightarrow  &  \left( \mathbf{1}, \mathbf{1,1}\right)&+ &  \left( \mathbf{1},\mathbf{1,5}\right).
\end{array}
\end{equation}

Its $\mathcal{N}=1$ twin $[1,2]\oplus \left( \mathbf{5+1}\right) [1,1]$ is
obtained by retaining the same bosonic sector, but truncating to the
complementary fermionic sector, namely retaining only the fermions
transforming as the $\mathbf{2}$ of $\Sp(1)$ :%
\begin{equation}\label{Sp1trunc2}
\begin{array}{lllllllllll}
\left( \mathbf{5},\mathbf{1}\right)  & \rightarrow  & \left( \mathbf{5},
\mathbf{1,1}\right)  &  &  \\
\left( \mathbf{4},\mathbf{6}\right)  & \rightarrow  &   \left( \mathbf{4}, \mathbf{2,1}\right)    \\
\left( \mathbf{3},\mathbf{14}+\mathbf{1}\right)  & \rightarrow & \left( \mathbf{3}, \mathbf{1,1}\right)   &+ &  \left( \mathbf{3}, \mathbf{1,5+1}\right)  \\
\left( \mathbf{2},\mathbf{14}^{\prime }+\mathbf{6}\right)  & \rightarrow  &    &+ &    \left( \mathbf{2}, \mathbf{2, 5+1}\right)   \\
\left( \mathbf{1},\mathbf{14}\right)  & \rightarrow  & &+ &  \left( \mathbf{1},\mathbf{1,5+1}\right).
\end{array}
\end{equation}

To summerise,  the $\mathcal{N}=2$ twin of the $\left\{
2,1\right\} $ pair is given by decomposing  $\left[ 3,2\right]$ into $
\mathcal{N}=2$ multiplets
\begin{equation}
\left[ 3,2\right] =\left[ 2,2\right] \oplus \mathbf{2}\left[ 2,3/2\right]
\oplus \left[ 2,1\right] ,
\end{equation}
and truncate out the $\Sp(2)$-doublet $\mathbf{2}$ of long, massive spin-3/2
multiplets. On the other hand, in order to get the $\mathcal{N}=1$ twin of
the $\left\{ 2,1\right\} $ pair, one  decomposes $\left[ 3,2%
\right] $ into $\mathcal{N}=1$ multiplets%
\begin{equation}
\left[ 3,2\right] =\left[ 1,2\right] \oplus \mathbf{4}\left[ 1,3/2\right]
\oplus \left( \mathbf{5}+\mathbf{1}\right) \left[ 1,1\right] \oplus \mathbf{4%
}\left[ 1,1/2\right] ,
\end{equation}%
and truncates out the $\mathbf{4}$ $\Sp(2)$-representations of long, massive
spin-3/2 and spin-1/2 multiplets.

\subsubsection{The $\N=2$ big twin} 

 Let us now introduce the big $\N_l=2$ twin S-fold operator $\s21b:=s_k\circ r_k$. The R-symmetry action in this case is given by, 
\be
r_k: \left\{\begin{array}{rrrrrrrrrrrrrr} Q_{i} &\mapsto& e^{-\frac{i\pi}{k}} Q_{i}, &\quad& Q_{3} &\mapsto& e^{\frac{i2\pi}{k}} Q_{3}  &\quad& Q_{4} &\mapsto&Q_{4}  \\[8pt]
\bar{Q}^{i} &\mapsto& e^{\frac{i\pi}{k}}  \bar{Q}^{i},&\quad& \bar{Q}^{3} &\mapsto& e^{-\frac{i2\pi}{k}}  \bar{Q}^{3} &\quad& \bar{Q}^{4} &\mapsto&  \bar{Q}^{4}
\end{array}\right.
\ee
where $i=1,2$. The S-duality operator is  given, as for the $\{3,1\}$ big twin, by,
\be\label{sk}
s_k: \left\{\begin{array}{rrrrrrrrr} Q_{A} &\mapsto& e^{\frac{i\pi}{k}} Q_{A};  \\[8pt]
\bar{Q}^{A} &\mapsto& e^{-\frac{i\pi}{k}}  \bar{Q}^{A}. 
\end{array}\right.
\ee
Hence, the composite action is given by
\be\label{b21twin}
\s21b:\left\{\begin{array}{rrrrrrrrrrrrrr} Q_{i} &\mapsto&  Q_{i}, &\quad& Q_{3} &\mapsto& e^{\frac{i3\pi}{k}} Q_{3}  &\quad& Q_{4} &\mapsto& e^{\frac{i\pi}{k}} Q_{4}  \\[8pt]
\bar{Q}^{i} &\mapsto&  \bar{Q}^{i},&\quad& \bar{Q}^{3} &\mapsto& e^{-\frac{i3\pi}{k}}  \bar{Q}^{3} &\quad& \bar{Q}^{4} &\mapsto&  e^{-\frac{i\pi}{k}}  \bar{Q}^{4}
\end{array}\right.\ee
 The R-symmetry is broken to $\SU(2)\times \Un(1)_R\times \Un(1)_F$ by $r_k$, where the first two factors make up the $\N=2$ R-symmetry $\SU(2)\times \Un(1)_R$ of the preserved $\N=2$ superalgebra.  The entire $\N=4$ vector multiplet transforms non-trivially under $\s21b$, as summarised in \autoref{N21breps}. The remaining $\Un(1)_F$ would seem to be spurious since, from \eqref{Sp1trunc}, we know that the maximal global symmetry of the big $\{2,1\}$ twin is $\SU(2)\times \Un(1)_R\subset \Sp(2)$. 
However, the $\s21b$-invariant sector is uncharged under the extra $\Un(1)$ so that the non-trivial global symmetry is indeed $\SU(2)\times \Un(1)_R$, as expected. 
\begin{table}
\[
\begin{array}{ccccccccc}
\hline
\hline
					&&\SU(2)		&&\Un(1)_R			&&\Un(1)_F			&& \s21b\\[4pt]
\hline
F^+		&&\rep{1}		&&0					&& 0					&&2\\[4pt]
\hline
\lambda^{i}	&&\rep{2}		&&1					&&0					&&2 \\[4pt]
\lambda^{3}	&&\rep{1}		&&\hspace{-0.3cm}-1	&&1					&&\hspace{-0.3cm}-1\\[4pt]
\lambda^{4}	&&\rep{1}		&&\hspace{-0.3cm}-1						&&\hspace{-0.3cm}-1	&&1\\[4pt]
\hline
\phi^{ij}				&&\rep{{1}}	&&2					&&0					&&2\\[4pt]
\phi^{i3}				&&\rep{2}		&&0					&&1					&&\hspace{-0.3cm}-1\\[4pt]
\phi^{i4}				&&\rep{2}		&&0					&&\hspace{-0.3cm}-1	&&1\\[4pt]
\phi^{34}				&&\rep{1}		&&\hspace{-0.3cm}-2	&&0					&&\hspace{-0.3cm}-2\\[4pt]
\hline
\hline
\end{array}
\]
\caption{The charges carried by the component fields of the $\N=4$ super Yang-Mills multiplet under the $\s21b$ S-fold operator (in units of $\pi/k$) and the invariant $\SU(2)\times \Un(1)_R\times \Un(1)_F\subset \SU(4)$  flavour/R-symmetry subgroup.}\label{N21breps}
\end{table}

Using the $\s21b$ charges it is straightforward to deduce the quadratic $\s21b$-invariant operators and, through their $\SU(2)\times\Un(1)_R$ representations, to collect them into massive long $\N=2$ supermultiplets. In term of the on-shell superfields of \cite{Siegel:1980bp, Howe:1981qj} we obtain one spin-2 and  one spin-1 supercurrent,
\be
J  = V \bar{V}, \quad J_{i}{}^{j} = V_i\bar{V}^j-\frac{1}{2}\delta_{i}{}^{j}V_k \bar{V}^k,
\ee
where $V$ is the $\N=2$ spin-1 on-shell superfield and $V_i$ is the  $\N=2$ spin-1/2 on-shell superfield \cite{Howe:1981qj}, transforming as a doublet  of the global symmetry $\Un(2)$. Hence, the $\s21b$ S-folding reproduces precisely the truncation \eqref{Sp1trunc} giving the big $\{2,1\}$ twin. The off-shell component field projection is given in   \autoref{field-proj}.

\subsubsection{The $\N=1$ S-fold construction} 

Let us now introduce the $\{2,1\}$ little twin S-fold operator $\s21l:=s_k\circ r_k$. As for the $\{3,1\}$ example, the R-symmetry action for the little is the same as that for the big twin, 
\be
r_k: \left\{\begin{array}{rrrrrrrrrrrrrr} Q_{i} &\mapsto& e^{-\frac{i\pi}{k}} Q_{i}, &\quad& Q_{3} &\mapsto& e^{\frac{i2\pi}{k}} Q_{3}  &\quad& Q_{4} &\mapsto&Q_{4}  \\[8pt]
\bar{Q}^{i} &\mapsto& e^{\frac{i\pi}{k}}  \bar{Q}^{i},&\quad& \bar{Q}^{3} &\mapsto& e^{-\frac{i2\pi}{k}}  \bar{Q}^{3} &\quad& \bar{Q}^{4} &\mapsto&  \bar{Q}^{4}
\end{array}\right.
\ee
where $i=1,2$. The difference again lies solely in the  S-duality operator,
\be\label{sk}
s_k: \left\{\begin{array}{rrrrrrrrr} Q_{A} &\mapsto& e^{-\frac{2i\pi}{k}} Q_{A};  \\[8pt]
\bar{Q}^{A} &\mapsto& e^{\frac{2i\pi}{k}}  \bar{Q}^{A}. 
\end{array}\right.
\ee
Comparing with the $\{3,1\}$ case, we note that (i) the S-duality phase is $\exp[\pm i\pi/k]$ for all big twins while (ii) the R-symmetry operator is the same for each pair of big and little twins, and  (iii) if $\N_b=n$ the S-duality on the corresponding little twin is given by $\exp[\mp in\pi/k]$. Note: (i)  simply reflects the fact that the supercharges transform uniformly under S-duality, so any change amounts to  a trivial redefinition of the S-fold; (ii) follows from the requirement that each twin pair has the same global symmetry, which is determined by the subalgebra commuting  with $r_k$ alone; (iii) is a consequence of breaking the $\N=4$ R-symmetry to $\N=\N_b$, which implies a single supercharge carries charge $\N_b$ and so can always be chosen to  be the $\N_l=1$ supercharge.

Hence, the composite action for the little $\{2,1\}$ twin is given by
\be\label{l21twin}
\s21l:\left\{\begin{array}{rrrrrrrrrrrrrr} Q_{i} &\mapsto&  e^{-\frac{3i\pi}{k}} Q_{i}, &\quad& Q_{3} &\mapsto& Q_{3}  &\quad& Q_{4} &\mapsto& e^{-\frac{2i\pi}{k}} Q_{4}  \\[8pt]
\bar{Q}^{i} &\mapsto& e^{\frac{3i\pi}{k}}  \bar{Q}^{i},&\quad& \bar{Q}^{3} &\mapsto&  \bar{Q}^{3} &\quad& \bar{Q}^{4} &\mapsto&  e^{\frac{2i\pi}{k}}  \bar{Q}^{4}
\end{array}\right.\ee
  Now only four supercharges survive, leaving an $\N=1$ superalgebra. 
 \begin{table}
\[
\begin{array}{ccccccccc}
\hline
\hline
					&&\SU(2)		&&\Un(1)_R			&&\Un(1)_F			&& \s21l\\[4pt]
\hline
F^+		&&\rep{1}		&&0					&& 0					&&\hspace{-0.3cm}-4\\[4pt]
\hline
\lambda^{i}	&&\rep{2}		&&1					&&0					&&\hspace{-0.3cm}-1 \\[4pt]
\lambda^{3}	&&\rep{1}		&&\hspace{-0.3cm}-1	&&1					&&\hspace{-0.3cm}-4\\[4pt]
\lambda^{4}	&&\rep{1}		&&\hspace{-0.3cm}-1						&&\hspace{-0.3cm}-1	&&\hspace{-0.3cm}-2\\[4pt]
\hline
\phi^{ij}				&&\rep{{1}}	&&2					&&0					&&2\\[4pt]
\phi^{i3}				&&\rep{2}		&&0					&&1					&&\hspace{-0.3cm}-1\\[4pt]
\phi^{i4}				&&\rep{2}		&&0					&&\hspace{-0.3cm}-1	&&1\\[4pt]
\phi^{34}				&&\rep{1}		&&\hspace{-0.3cm}-2	&&0					&&\hspace{-0.3cm}-2\\[4pt]
\hline
\hline
\end{array}
\]
\caption{The charges carried by the component fields of the $\N=4$ super Yang-Mills multiplet under the $\s21l$ S-fold operator (in units of $\pi/k$) and the invariant $\SU(2)_F\times \Un(1)_R\times \Un(1)_F\subset \SU(4)$  flavour/R-symmetry subgroup. }\label{N21lreps}
\end{table}
Excluding $k=2$ the entire $\N=4$ vector multiplet transforms non-trivially under $\s21l$, as summarised in \autoref{N21lreps}. Specialising to $k=3$ it is straightforward to deduce the quadratic $\s21l$-invariant operators and, through their $\Un(2)_F\times\Un(1)_R$ representations, to collect them into massive long $\N=1$ supermultiplets, yielding one spin-2 and six spin-1 multiplets in the $\rep{5}+\rep{1}$ of $\Sp(2)$ as required. See  \autoref{field-proj}.

\subsubsection{Dual supergravity theories}

Ungauged $D=5$, $\mathcal{N}_{b}=4$ supergravity coupled to one vector
multiplet has a {twin} given by the $\mathcal{N}_{l}=2$ supergravity
coupled to six vector multiplets, based on the semi-simple rank-3 Jordan
algebra $\mathbb{R}\oplus \mathbf{\Gamma }_{1,5}$ \cite{Roest:2009sn}. The
bosonic sectors of such twins are determined by the common scalar symmetric
coset%
\begin{equation}
\SO(1,1)\times \frac{\SO\left( 1,5\right) }{\SO(5)},\label{D=5}
\end{equation}%
where $\SO(1,1)\times \SO(1,5)$ is the reduced structure group of $\mathds{R}%
\oplus \mathbf{\Gamma }_{1,5}$. This is the $D=5$ analog of the $D=4$ $%
\left\{ 4,2\right\} $ supergravity twin pair \cite{Ferrara:2008ap}, with
common symmetric coset%
\begin{equation}
\frac{\SL(2,\mathbb{R})}{\Un(1)}\times \frac{\SO\left( 2,6\right) }{\SO(2)\times
\SO(6)},
\end{equation}%
which is the R-map image of \eqref{D=5}.

Following the discussion of \autoref{31dual}, together with the $\s21b/\s21l$ S-foldings and the observation that both the $\{2,1\}$ W-SCFT and the  $\{4,2\}$ Poincar\'e supergravity twins are truncations of the $\{3,1\}$ W-SCFT and the  $\{6,2\}$ Poincar\'e supergravity twins, respectively, we would anticipate analogous `twin' truncations yielding the candidate bulk dual   $\{4,2\}$  gauged twin  supergravities.  The big gauged $\N=4$ twin corresponds to a further consistent truncation of the special case described in \cite{Gunaydin:1985cu}, in which  $\SU(4)$ gauged $\N=8$ supergravity is truncated down to  Romans' gauged $\N=4$ supergravity \cite{Romans:1985ps} coupled to a single vector multiplet. Note, in Romans' gauged $\N=4$ supergravity the  vectors of $\N=4$ Poincar\'e  
 supergravity sitting in the $\rep{5}$ of $\Sp(2)$  are replaced by three vectors and two ``self-dual'' two-forms in the $\rep{3}_0$ and $\rep{1}_2+\rep{1}_{-2}$ of $\Un(2)\subset\Sp(2)$, as required by the preservation of supersymmetry, which  is precisely reflected by the multiplet structure of the candidate dual $\N=2$ W-SCFT, cf.~\autoref{field-proj}. Similarly, the little $\N=2$ twin corresponds to a further consistent truncation of the  $\N=8\rightarrow\N=2$ case given in \cite{Gunaydin:1985cu} (and described in \autoref{31dual}) down to $\Un(1)_R\times \Un(2)$ gauged $\N=2$ supergravity  coupled to $3+1$ vector multiplets and $1+1$ ``self-dual'' tensors multiplets in the $\rep{3}^{0}_{0}+\rep{1}^{0}_{0}$ and $\rep{1}^{0}_{2}+\rep{1}^{0}_{-2}$, as required by supersymmetry \cite{Gunaydin:1984qu, Gunaydin:1999zx}, again reflecting the structure of the candidate dual $\N=1$ W-SCFT. See \autoref{field-proj}. Note, the $\Un(1)_R$ gauge factor is required for an AdS vacuum \cite{Gunaydin:1999zx, Gunaydin:2000xk}. As for the $\{6,2\}$ case discussed in \autoref{31dual}, we do not anticipate that these truncations can be obtained using purely (conventional) geometric symmetries, but require instead  the twin $\s21b/\s21l$ S-foldings implemented on $\SU(4)$ gauged $\N=8$ supergravity. Note, the need to invoke S-duality here implies that they are intrinsically non-perturbative. We would also expect to be able to obtain the gauged twin supergravities  directly through twin gaugings of the  $\{4,2\}$ Poincar\'e supergravities following \cite{Gunaydin:1999zx, DallAgata:2001wgl}.

\subsubsection{Double-copy construction}

The ``parent'' $[3, 2]$ multiplet is given by 
\begin{equation}
\left[ 2,1\right] ^{A}\otimes \left[ 1,1\right] ^{\tilde{A}}=\left[ 3,2\right].
\end{equation}%
To obtain the $\mathcal{N}=2$ twin of the $\left\{ 2,1\right\} $ pair, first
decompose the Left factor into $\mathcal{N}=1$ multiplets :
\begin{equation}
\left( \left[ 1,1\right] ^{A}\oplus 2\left[ 1,1/2\right] ^{A}\right) \otimes %
\left[ 1,1\right] ^{\tilde{A}}=\left[ 3,2\right] ,
\end{equation}%
and then truncate the Left factor to the $\sigma $-invariant sector. By
using the rule that adjoint and non-adjoint representations do not talk to
one another in the double copy, one obtains%
\begin{equation}
\left( \left[ 1,1\right] ^{A_{0}}\oplus 2\left[ 1,1/2\right] ^{\rho }\right)
\otimes \left[ 1,1\right] ^{\tilde{A}}=\left[ 2,2\right] \oplus \left[ 2,1%
\right] .
\end{equation}%
To obtain the $\mathcal{N}=1$ twin of the $\left\{ 2,1\right\} $ pair, one
has then to decompose the Right factor into $\mathcal{N}=0$ multiplets and
truncate to the $\tilde{\sigma}$-invariant sector :%
\begin{equation}
\left( \left[ 1,1\right] ^{A_{0}}\oplus 2\left[ 1,1/2\right] ^{\rho }\right)
\otimes \left( \left[ 0,1\right] ^{\tilde{A}_{0}}+2\left[ 0,1/2\right] ^{%
\tilde{\rho}}+\left[ 0,0\right] ^{\tilde{A}_{0}}\right) =\left[ 1,2\right]
\oplus 6\left[ 1,1\right] ,
\end{equation}%
where $6$ must be specified as $\mathbf{5+1}$ under $\Sp(2)$.

\section{Twin W-supergravities}\label{Wsugra}

In \cite{Ferrara:2018iko} it was argued that W-supergravities, which possess a  spin-4 field in place of the conventional graviton, follow from the effective field theory limit of asymmertric S-foldings of   string theory. For recent developments on W-supergravities see \cite{Ferrara:2018wqd, Ferrara:2018wlb}. The string S-fold is effected by a T-duality twist and an S-duality twist combined with a new G-duality twist, which ensures that the S-fold only acts on the right movers, and a H-duality twist, which ensures that the S-fold is Lorentz covariant. The combined S-G-H-duality is an automorphism of the string theory, so the S-fold is a \emph{bona fide} projection. The string S-fold reproduces the field theory S-fold on the  right-moving sector and so  the spectrum can be calculated using the product or ``double-copy'' of the corresponding W-SCFTs, where level matching forbids   products amongst the right W-SCFT and the conventional massless states still present in the left-moving sector. 

For example, the spectrum of $\N=7$ W-supergravity \cite{Ferrara:2018iko}  follows from the product between the $\N=4$ left-moving sector and the S-folded $\N=3$ right-moving sector of  type II strings on $T^6$, which at the lowest level reduces to
\be
[4, 2]_L\times [3,2]_R = [7,4].
\ee
By considering various degrees of supersymmetry in the factors we can construct the spectra of the would-be W-supergravities \cite{Ferrara:2018iko},  with all $0\leq \N\leq 7$ as summarised in \autoref{tab:spin2}. We can also generate (almost) arbitrary  ``matter'' couplings  for $\N\leq 6$. Note, although we have included the $[8,4]$ multiplet for completeness, it does not correspond to any W-supergravity as the S-fold always breaks some supersymmetry (at least not without some further, as yet to be determined, novel ingredients). It is also useful in that it provides a ``parent'' multiplet for the first example of twin W-supergravities. 

Using \autoref{tab:spin2} and \autoref{tab:moreproducts}, together with the branchings given in \autoref{tab:branchings}, it is then straightforward to follow \cite{Anastasiou:2016csv} (cf.~\autoref{dctwin}) to construct would-be twin W-supergravities, which have identical bosonic symmetries and spectra.  To go beyond this would require a better understanding of the S-folded vertex operators, which we leave for future work. Let us give an example, generalising the prototypical case of the  $\{6,2\}$ twins in conventional supergravity. As indicated, the maximally supersymmetric ``parent'' multiplet is given by 
\be
[8, 4]_{\text{parent}}=[4,2]\times [4,2],
\ee
which corresponds to the parent $\N=8$ supergravity of the $\{6,2\}$ twin supergravities, although does not exist itself as a W-supergravity. 

The big twin is given by  branching the right theory down to $\N=2$  with $\Sp(2)_{\tilde{R}}\times\Sp(2)_{\tilde{F}}\subset\Sp(4)_{\tilde{R}}$,
\be
[4,2]\times\left([2,2]+ \rep{4}[2, 3/2]+\rep{5}[2,1]\right)=[4,2]\times [2,2]+ \rep{5} [4,2]\times [2,1]=[6, 4]+\rep{5}[6,3],
\ee
where we have employed the rule that integer and half-integer multiplets to not talk to-one-another in the product, as described in \autoref{dctwin}. This reflects the property of the scattering amplitude double-copy that adjoint and non-adjoint representations of the gauge group do not mix \cite{Johansson:2014zca}.  The multiplicities are given as representations of $\Sp(2)_{\tilde{F}}$ and the total global symmetry is $\Sp(6)_R\times \Sp(2)_{\tilde{F}}$. 

Following \cite{Anastasiou:2016csv} the little  twin is given by further branching the left theory down to $\N=0$ with $\Sp(4)_{F}\equiv\Sp(4)_R$,
\be
\left([0,2]+\rep{8}[0, 3/2]+\rep{27}[0,1]+\rep{48}[0, 1/2]+\rep{42}[0,0]\right)\times\left( [2,2]+ \rep{4}[2, 3/2]+\rep{5}[2,1]\right),
\ee
which yields,
\be
\begin{array}{llllll}
[2,4]
&+(\rep{(1,1)+(1,5)+(27,1)+(8,4)})[2,3]\\
&+(\rep{(1,1)+(27,5)+(27,1)+(8,4)+(48,4)+(42,1)})[2,2]\\
&+(\rep{(27,1)+(48,4)+(42,5)})[2,1]\\
\end{array}
\ee
where we have given the multiplicities as $\Sp(4)_F\times\Sp(2)_{\tilde{F}}$ 
representations. These may be collected into irreducible $\Sp(6)_F$ representations,
\be
[2,4]+\rep{65}[2,3]+\rep{429}[2,2]+\rep{429'}[2,1],
\ee
so that the total global symmetry is $\Sp(6)_F\times \Sp(2)_{\tilde{R}}$. We see that the big and little twins thus have the same global symmetries. 
The bosonic spectra match. For instance, the big twin has $71$ spin-3 states in the $\rep{(65+1,1)+(1,5)}$ of $\Sp(6)_R\times\Sp(2)_{\tilde{F}}$, while the little twin has $71$ spin-3 states in the $\rep{(1,5+1)}+\rep{(65,1)}$ of  $\Sp(6)_F\times \Sp(2)_{\tilde{R}}$. Similarly, the spin-2 states sit in the  $\rep{(429+65+1,1)+(65,5)}$ and $\rep{(1,1)+(65, 5+1)+(429,1)}$ of  $\Sp(6)_R\times\Sp(2)_{\tilde{F}}$ and $\Sp(6)_F\times \Sp(2)_{\tilde{R}}$, respectively.

Using the same methodology we  obtain the   $\{5,1\}, \{4,2\}, \{3,1\}$ and $\{2,1\}$ twin W-supergravities, analogous to the conventional $D=4$ twins of \autoref{PYR}. There may also be further $D=4$ twins, since we also have parents with $\N=3$ factors, as well as twins in other dimensions, as suggested by \autoref{PYR}, but we leave the complete classification for future work.

  \begin{table}[ht]
 \begin{center}
%\begin{ruledtabular}
\begin{tabular}{|ccccl|}
\hline
\hline
  $[4,2]$ &$\otimes$  & $[4,2] $&$=$ &$[8,4]$\\ [5pt]
  $[4,2]$ &$\otimes$  & $[3,2] $&$=$ & $[7,4]$\\ [5pt]
  $[4,2]$ &$\otimes$  & $[2,2] $&$=$ & $[6,4]$\\ [5pt]
  $[4,2]$ &$\otimes$  & $[1,2] $&$=$ & $[5,4]$\\ [5pt]
  $[4,2]$ &$\otimes$  & $[0,2] $&$=$ & $[4,4]$\\ [5pt]
  \hline
  $[3,2]$ &$\otimes$  & $[3,2] $&$=$ & $[6,4]+ [6,3]$\\ [5pt]
  $[3,2]$ &$\otimes$  & $[2,2] $&$=$ & $[5,4] + [5,3] $\\ [5pt]
  $[3,2]$ &$\otimes$  & $[1,2] $&$=$ & $[4,4]+ [4,3] $\\ [5pt]
  $[3,2]$ &$\otimes$  & $[0,2] $&$=$ & $[3,4]+ [3,3]$\\ [5pt]
  \hline
$[2,2]$ &$\otimes$  & $[2,2] $&$=$ & $[4,4]+ [4,3] + [4,2]$\\ [5pt]
  $[2,2]$ &$\otimes$  & $[1,2] $&$=$ & $[3,4]+ [3,3] + [3,2] $\\ [5pt]
  $[2,2]$ &$\otimes$  & $[0,2] $&$=$ & $[2,4]+ [2,3] + [2,2]$\\ [5pt]
\hline
  $[1,2]$ &$\otimes$  & $[1,2] $&$=$ & $[2,4]+ [2,3] + [2,2] + [2,1]$\\ [5pt]
  $[1,2]$ &$\otimes$  & $[0,2] $&$=$ & $[1,4]+ [1,3] + [1,2] + [1,1]$\\ [5pt]
  \hline
    $[0,2]$ &$\otimes$  & $[0,2] $&$=$ & $[0,4] + [0,3] +[0, 2]+ [0, 1]+  [0, 0]$\\ [5pt]

\hline
   \hline
\end{tabular}
\caption{Products of massive spin-2 multiplets. \label{tab:spin2}}
%\end{ruledtabular}
 \end{center}
\end{table}

  \begin{table}[ht]
 \begin{center}
%\begin{ruledtabular}
\begin{tabular}{|ccccl|}
\hline
\hline
      $[4,2]$ &$\otimes$  & $[3,3/2] $&$=$ & $[7,7/2]$\\ [5pt]
    $[4,2]$ &$\otimes$  & $[2,3/2] $&$=$ & $[6,7/2]$\\ [5pt]
    $[4,2]$ &$\otimes$  & $[2,1] $&$=$ & $[6,3]$\\ [5pt]
  \hline
    $[3,2]$ &$\otimes$  & $[0,3/2] $&$=$ & $[3,7/2]+ [3,5/2]$\\ [5pt]
    $[3,2]$ &$\otimes$  & $[0,1] $&$=$ & $[3,3]+ [3,2]$\\ [5pt]
  $[3,3/2]$ &$\otimes$  & $[0,3/2] $&$=$ & $[3,3]$\\ [5pt]
  $[3,3/2]$ &$\otimes$  & $[0,1/2] $&$=$ & $[3,2]$\\ [5pt]
  \hline
    $[2,2]$ &$\otimes$  & $[0,1] $&$=$ & $[2,3] + [2,2]+ [2,1]$\\ [5pt]
    $[2,3/2]$ &$\otimes$  & $[0,3/2] $&$=$ & $[2,3]+ [2,2]$\\ [5pt]
        $[2,3/2]$ &$\otimes$  & $[0,1/2] $&$=$ & $[2,2]+ [2,1]$\\ [5pt]\
        $[2,1]$ &$\otimes$  & $[0,2] $&$=$ & $[2,3]$\\ [5pt]
                $[2,1]$ &$\otimes$  & $[0,1] $&$=$ & $[2,2]$\\ [5pt]

\hline
   \hline
\end{tabular}
\caption{Some relevant products of massive long multiplets. \label{tab:moreproducts}}
%\end{ruledtabular}
 \end{center}
\end{table}

  \begin{table}[ht]
 \begin{center}
%\begin{ruledtabular}
\begin{tabular}{|lll|}
\hline
\hline
  $[4,2]$ &$\rightarrow$  & $[3,2]+ \rep{2}\times[3, 3/2]$\\ [5pt]
  $[4,2]$ &$\rightarrow$  & $[2,2]+ \rep{4}\times[2, 3/2]+ \rep{5}\times[2, 1]$ \\ [5pt]
  $[4,2]$ &$\rightarrow$  & $[1,2]+ \rep{6}\times[1, 3/2]+ \rep{14}\times[1, 1]+ \rep{14'}\times[1, 1/2]$\\ [5pt]
  $[4,2]$ &$\rightarrow$  & $[0,2]+ \rep{8}\times[0, 3/2]+ \rep{27}\times[0, 1]+ \rep{48}\times[0, 1/2]+ \rep{42}\times[0, 0]$\\[5pt]
   \hline
    $[3,2]$ &$\rightarrow$  & $[2,2]+ \rep{2}\times[2, 3/2]+ \rep{1}\times[2, 1]$ \\ [5pt]
  $[3,2]$ &$\rightarrow$  & $[1,2]+ \rep{4}\times[1, 3/2]+ (\rep{5+1})\times[1, 1]+ \rep{4}\times[1, 1/2]$\\ [5pt]
  $[3,2]$ &$\rightarrow$  & $[0,2]+ \rep{6}\times[0, 3/2]+ (\rep{14+1})\times[0, 1]+ \rep{14'+6}\times[0, 1/2]+ \rep{14}\times[0, 0]$\\[5pt]
  \hline
  $[2,2]$ &$\rightarrow$  & $[1,2]+ \rep{2}\times[1, 3/2]+ \rep{1}\times[1, 1]$\\ [5pt]
  $[2,2]$ &$\rightarrow$  & $[0,2]+ \rep{4}\times[0, 3/2]+ (\rep{5+1})\times[0, 1]+ \rep{4}\times[0, 1/2]+ \rep{1}\times[0, 0]$\\[5pt]
\hline
  $[1,2]$ &$\rightarrow$  & $[0,2]+ \rep{2}\times[0, 3/2]+ \rep{1}\times[0, 1]$\\[5pt]
\hline
   \hline
\end{tabular}
\caption{Branchings of massive spin-2 multiplets under $\Sp(\N')\times \Sp(\N-\N')\subset\Sp(\N)$. Multiplicities are given in terms of $\Sp(\N-\N')$ representations. \label{tab:branchings}}
%\end{ruledtabular}
 \end{center}
\end{table}

\section{Conclusions}

We have argued that there are twin W-SCFTs using S-folds preserving $\N<3$ supersymmetry. The lowest level spectra may be deduced from the ``double-copy'' of massive long spin$\leq 1$ multiplets. Similarly, at the level of spectra and symmetries there exist twin W-supergravities. There are a number of directions we will consider in future work. Perhaps most obviously is the need, given their intrinsically non-perturbative nature, of a more complete understanding of the twins, and the W-SCFTs with $\N<3$ in general. In particular, a string/F-theory embedding   would  lend further support to their existence and twiness beyond spectra alone. One might also consider their central charges.  For instance, it is known (essentially using representation theory together with known properties of $\N=2$ theories alone) that the $\N=3$ theories obey $a=c$ \cite{Aharony:2015oyb}. This raises the possibility of relations (if any) amongst the central charges of the twins.   We will also generalise to other dimensions, as suggested by the twin pyramid \autoref{PYR}. The $D=3,4$ levels of \autoref{PYR} suggest the possibility of W-SCFTs in $D=2,3$. The $D=6$ layer, on the other hand, poses a puzzle as the unique $D=5$ superconformal group obstructs the existence of W-SCFT twins with distinct degrees of supersymmetry. The W-supergravities raise similar questions, especially with regard to their twinness beyond spectra/symmetries and further examples in $D=3,5,6$.

\acknowledgments
We are grateful to Sara Tahery for posing the question during the Erice International Subnuclear Physics Summer School that led us to this work:  are there twin conformal field theories? 
We are grateful for stimulating conversations with Sergio Ferrara on W-supergravities. 
We are grateful   to Philip Candelas for hospitality at the Mathematical Institute, University of Oxford, where part of this work was completed. MJD is grateful   to  Marlan Scully for his hospitality in the Institute for Quantum Science and Engineering, Texas A\&M University. MD acknowledges the Leverhulme Trust for an Emeritus Fellowship and the Hagler Institute for Advanced Study at Texas A\&M for a Faculty Fellowship.  The work of LB is supported by a Schr\"odinger Fellowship.  The work of MJD is supported by was  supported in part by the STFC under rolling grant ST/P000762/1. 

\appendix

\section{Supercurrent component projection}\label{field-proj}

In terms of component fields the $\N=4$ supercurrent $J_{AB, CD}$ is given by \cite{Bergshoeff:1980is, Ferrara:1998ej},
\be
\begin{array}{ccccccccccccccccccccccc}
&&g_{\mu\nu}, 
&&\psi^{A}_{\mu}
&&A_\mu{}^{A}{}_{B} && A_{\mu\nu}^{AB} 
&&\chi^{A} &&  \chi_{C}^{AB}
&&\varphi &&  \varphi^{AB} &&  \varphi_{CD}^{AB},\\  
&& \rep{1}, 
&& \rep{4}+\rep{\overline{4}}&& \rep{15}&& \rep{6}_\C&& \rep{4}+\rep{\overline{4}}&& \rep{20}+\rep{\overline{20}}&& \rep{1}_\C&& \rep{10}+\rep{\overline{10}}&& \rep{20'}
\end{array}
\ee 
where we have indicated the corresponding $\SU(4)$ representations and 
\begin{subequations}\label{Jcomps}
\begin{eqnarray}
g_{\mu\nu} &= &\begin{array}{l}\frac{1}{2}\left(\eta_{\mu\nu} F^{-}_{\rho\sigma}F^{+ \rho\sigma}-4F^{-}_{\mu}{}^{\rho}F^{+}_{\nu\rho}+\text{h.c.}\right) - \frac{1}{2}\bar{\lambda}_{A}\gamma_{(\mu}\partial_{\nu)}^{\leftrightarrow}\lambda^A  \\[4pt] 
+\eta_{\mu\nu}\partial^\rho\phi^{AB}\partial_\rho \phi_{AB}-2\partial_\mu \phi^{AB} \partial_\nu \phi_{AB} -\frac{1}{3}\left(\eta_{\mu\nu}\Box - \partial_\mu\partial_\nu\right)\phi^{AB}\phi_{AB} \end{array}\\
{\psi}_{\mu}^{A}   &=& -(\sigma F^-)\gamma_\mu{\lambda}^A + 2i \phi^{AB}\partial_{\mu}^{\leftrightarrow}{\lambda}_B+\frac{4}{3}i\sigma_{\mu\rho}\partial^\rho(\phi^{AB}{\lambda}_B)\\
A_\mu{}_{A}{}^{B} &=& \phi_{AC}\partial_{\mu}^{\leftrightarrow}\phi^{CB}+\bar{\lambda}_{A}\gamma_{\mu}\lambda^{B} - \frac{1}{4}\delta_{A}{}^{B}\bar{\lambda}_{C}\gamma_{\mu}\lambda^{C}\\
A_{\mu\nu}^{AB} &=&\bar{\lambda}^{A}\sigma_{\mu\nu}{\lambda}^{B}+2i\phi^{AB}F^{+}_{\mu\nu}\\
\chi^{A} &=& \sigma F^{+} \lambda^A\\
\chi^{AB}_{E} &=& \frac{1}{2}\epsilon^{ABCD}\left(\phi_{CD}\lambda_E +\phi_{CE}\lambda_{D} \right)\\
\varphi &=&  F^{-}_{\mu\nu}F^{-\mu\nu}\\
\varphi^{AB} &=&  \bar{\lambda}^A \lambda^B\\
 \varphi_{CD}^{AB} &=&\phi^{AB}\phi_{CD} - \frac{1}{12}\delta_{C}{}^{[A}\delta_{D}{}^{B]}\phi^{EF}\phi_{EF} 
\end{eqnarray}
\end{subequations}

\paragraph{The $\{3,1\}$ twins:}

The fields  transform under the $\s31b$ S-fold \eqref{b31twin} with weights (in units of $2\pi/k$),
\be
\begin{array}{ccccccccccccccccc}
&&\phi^{ab} &&\phi^{a4} &&  \lambda^{4} &&\lambda^{a} &&  F^{+}\\  
&& 1&& -1&& -1 && 1&& 1 
\end{array}
\ee  
which project \eqref{Jcomps}  onto  a single spin-2 $\N=3$ supercurrent, with component field schematically given by 
\be
\begin{array}{llllllllllllllllllllllllllllll}
(g_{\mu\nu},   && {\psi}_{\mu}^{a} & {\psi}_{\mu}{}_{a},  && A_\mu{}_{a}{}^{b} & A_{\mu\nu}^{a4} &A_{\mu\nu}{}_{a4} & A_\mu{}_{4}{}^{4}, && \chi^{4} & \chi_{4} & \chi^{ab}_{4} & \chi_{ab}^{4} && \chi^{a4}_{4} & \chi_{a4}^{4},  && \varphi^{a4} & \varphi_{a4} & \varphi^{a4}_{b4}) 
\end{array}
\ee
carrying $\Un(3)$ representations given in \eqref{31breps}. For the complete characterisation of the $\N=3$ Weyl supercurrent-multiplet see \cite{vanMuiden:2017qsh}.

For $k=4$ the fields  transform under the $\s31l$ S-fold \eqref{l31twin} with weights (in units of $\pi/2$),
\be
\begin{array}{ccccccccccccccccc}
&&\phi^{ab} &&\phi^{a4} &&  \lambda^{4} &&\lambda^{a} &&  F^{+}\\  
&& -1&& 1&& -1 && 1&& -1 
\end{array}
\ee  
which project \eqref{Jcomps}  onto  a single spin-2 $\N=1$ supercurrent 
\be
\begin{array}{llllllllllllllllllll}
(g_{\mu\nu},   && {\psi}_{\mu}^{4}, &&  {\psi}_{\mu}{}_{4},  && A_\mu{}_{4}{}^{4}) 
\end{array}
\ee
and 14 spin-1 $\N=1$ supercurrents 
\be
\begin{array}{llllllllllllllllllll}
(A_\mu{}_{a}{}^{b}, && \chi^{a4}_{b}, && \varphi_{a4}^{b4}), && (A_{\mu\nu}{}^{a4}, && {\chi}^{a}, && \varphi^{a4}),  && (A_{\mu\nu}{}_{b4}  && {\chi}_{b},        && \varphi_{b4}),   
\end{array}
\ee
 transforming in the $\rep{8}_0, \rep{3}_2,  \rep{\bar{3}}_{-2}$ of $\Un(3)$, respectively, as can be checked directly using the supersymmetry transformation rules of the $\N=4$ super Yang-Mills multiplet with the variational parameter $\varepsilon^A$ restricted to $\varepsilon^4$.

\paragraph{The $\{2,1\}$ twins:}
 The fields  transform under the big twin  $\s21b$ S-fold \eqref{b21twin} with weights (in units of $\pi/k$),
\be
\begin{array}{ccccccccccccccccc}
&&\phi^{34} &&\phi^{i4}&&\phi^{i3} &&\phi^{ij} &&  \lambda^{4}  &&  \lambda^{3} &&\lambda^{i} &&  F^{+}\\  
&& -2&& 1&& -1 && 2&& 1 &&-1 && 2 &&2
\end{array}
\ee  
which project \eqref{Jcomps} onto  a single spin-2 and a single spin-1 $\N=2$ supercurrent, given schematically by 
\be
\begin{array}{llllllllllllllllllllllllllllll}
(g_{\mu\nu},   && {\psi}_{\mu}^{i} & {\psi}_{\mu}{}_{i},  && A_\mu{}_{i}{}^{j} & A_{\mu\nu}^{34} &A_{\mu\nu}{}_{34} & A_\mu{}_{3}{}^{3}+A_\mu{}_{4}{}^{4}, &&  \chi^{i3}_{3}+  \chi^{i4}_{4} & \chi_{i3}^{3}+  \chi_{i4}^{4}, &&  \varphi^{34}_{34}) 
\end{array}
\ee
and
\be
\begin{array}{llllllllllllllllllll}
( A_\mu{}_{3}{}^{3}-A_\mu{}_{4}{}^{4}, && \chi^{i3}_{3}-  \chi^{i4}_{4} & \chi_{i3}^{3}- \chi_{i4}^{4}, && \varphi_{i3}^{j3}+ \varphi_{i4}^{j4} & \varphi_{ij} & \varphi^{ij}).\end{array}
\ee
where the spin 3/2, 1, 1/2 and 0  fields are in the $\rep{2}_1+\rep{2}_{-1}$,  $\rep{3}_0+\rep{1}_{2}+\rep{1}_{-2}+\rep{1}_{0}+\rep{1}_{0}$, $\rep{2}_1+\rep{2}_{-1}+\rep{2}_1+\rep{2}_{-1}$ and $\rep{3}_0+\rep{1}_{2}+\rep{1}_{-2}+\rep{1}_{0}$ of $\Un(2)_R$, respectively,  in agreement with the decomposition of \eqref{Sp1trunc} under $\Un(2)\in \Sp(2)$.   
The precise linear combinations are uniquely determined by closure under the supersymmetry transformations  given in \cite{Bergshoeff:1980is} with the variational parameter $\varepsilon^A$ restricted to $\varepsilon^i$.

For $k=3$ the fields  transform under the little twin $\s21l$ S-fold \eqref{l21twin} with weights (in units of $\pi/3$),
\be
\begin{array}{ccccccccccccccccc}
&&\phi^{34} &&\phi^{i4}&&\phi^{i3} &&\phi^{ij} &&  \lambda^{4}  &&  \lambda^{3} &&\lambda^{i} &&  F^{+}\\  
&& -2&& 1&& -1 && 2&& -2 &&2 && -1 &&2
\end{array}
\ee  
which project \eqref{Jcomps} onto  a single spin-2 $\N=1$ supercurrent, given schematically by  
\be
\begin{array}{llllllllllllllllllll}
(g_{\mu\nu},   && {\psi}_{\mu}^{3} & {\psi}_{\mu}{}_{3},   && A_\mu{}_{3}{}^{3}) 
\end{array}
\ee
and $5+1$ spin-1 $\N=1$ supercurrents
\be
\begin{array}{l}
(A_\mu{}_{i}{}^{j} \quad A_{\mu\nu}^{34} \quad A_{\mu\nu}{}_{34} \quad A_\mu{}_{3}{}^{3}-3A_\mu{}_{4}{}^{4}, \\ ~~\chi^{i3}_{j} \quad \chi_{j3}^{i} \quad\chi^{ij}_{3} \quad \chi_{ij}^{3}\quad\chi^{34}_{4} \quad\chi_{34}^{4}\quad \chi^{4} \quad \chi_{4} \\
~~ \varphi_{i3}^{j3}\quad\varphi^{34}\quad\varphi_{34} \quad \varphi_{34}^{34}), 
\end{array}
\ee
where the spin 3/2, 1, 1/2 and 0  fields are in the $\rep{1}^{1}_{0}+\rep{1}^{-1}_{0}$,  $\rep{3}^{0}_{0}+\rep{1}^{0}_{2}+\rep{1}^{0}_{-2}+\rep{1}^{0}_{0}+\rep{1}^{0}_{0}$, $\rep{3}^{1}_{0}+\rep{1}^{1}_{2}+\rep{1}^{1}_{-2}+\rep{1}^{1}_{0}+\text{c.c.}$ and $\rep{3}^{0}_{0}+\rep{1}^{0}_{2}+\rep{1}^{0}_{-2}+\rep{1}^{0}_{0}$ of $\Un(2)_F\times \Un(1)_R$, respectively,  in agreement with the decomposition of \eqref{Sp1trunc2} under $\Un(1)_R\times \Un(2)_F\in \Sp(1)\times\Sp(2)$.   
The precise linear combinations are uniquely determined by closure under the supersymmetry transformations  given in \cite{Bergshoeff:1980is} with the variational parameter $\varepsilon^A$ restricted to $\varepsilon^3$.

\providecommand{\href}[2]{#2}\begingroup\raggedright\endgroup

\end{document}